\documentclass[10pt,journal,compsoc]{IEEEtran}

\usepackage{algorithmic}
\usepackage{multirow}
\usepackage[pdftex]{graphicx}
\usepackage{url}
\usepackage{amsmath}
\usepackage{hyperref}
\usepackage{xcolor}
\usepackage{url}
\usepackage{array}
\usepackage{stfloats}
\usepackage[caption=false,font=footnotesize,labelfont=sf,textfont=sf]{subfig}
\usepackage[nocompress]{cite}

\begin{document}

\title{Temporal Upsampling of Depth Maps\\ Using a Hybrid Camera}

\author{Ming-Ze~Yuan,
Lin~Gao$^{*}$, 
Hongbo~Fu, 
and~Shihong~Xia$^{*}$
\IEEEcompsocitemizethanks{
\IEEEcompsocthanksitem Corresponding Author: Lin Gao (gaolin@ict.ac.cn) and Shihong Xia (xsh@ict.ac.cn)
\IEEEcompsocthanksitem M.-Z. Yuan is with the Beijing Key Laboratory of Mobile Computing and Pervasive Device, Institute of Computing Technology, Chinese Academy of Sciences, and also with the University of Chinese Academy of Sciences, Beijing, 100190, China.\protect\\
E-mail: yuanmingze@ict.ac.cn.\protect
\IEEEcompsocthanksitem L. Gao is with the Beijing Key Laboratory of Mobile Computing and Pervasive Device, Institute of Computing Technology, Chinese Academy of Sciences, Beijing, 100190, China.\protect\\  E-mail: gaolin@ict.ac.cn.\protect
\IEEEcompsocthanksitem H. Fu is with the School of Creative Media, City University of Hong Kong.\par E-mail: hongbofu@cityu.edu.hk
\IEEEcompsocthanksitem S. Xia is with the Beijing Key Laboratory of Mobile Computing and Pervasive Device, Institute of Computing Technology, Chinese Academy of Sciences, Beijing, 100190, China.\protect\\  E-mail: xsh@ict.ac.cn.\protect
}
}

\markboth{IEEE TRANSACTIONS ON VISUALIZATION AND COMPUTER GRAPHICS,~Vol.~xx, No.~xx, July~2017}%
{Shell \MakeLowercase{\textit{et al.}}: Bare Demo of IEEEtran.cls for Computer Society Journals}

\IEEEtitleabstractindextext{
\begin{abstract}
In recent years, consumer-level depth cameras have been adopted for various applications. However, they often produce depth maps at only a moderately high frame rate (approximately 30 frames per second), preventing them from being used for applications such as digitizing human performance involving fast motion. On the other hand, low-cost, high-frame-rate video cameras are available. This motivates us to develop a hybrid camera that consists of a high-frame-rate video camera and a low-frame-rate depth camera and to allow temporal interpolation of depth maps with the help of auxiliary color images. To achieve this, we develop a novel algorithm that reconstructs intermediate depth maps and estimates scene flow simultaneously. We test our algorithm on various examples involving fast, non-rigid motions of single or multiple objects. Our experiments show that our scene flow estimation method is more precise than a tracking-based method and the state-of-the-art techniques.
\end{abstract}

}

\maketitle

\IEEEpeerreviewmaketitle

\IEEEraisesectionheading{\section{Introduction}\label{sec:introduction}}
\IEEEPARstart{I}{}n recent years, low-cost depth cameras such as Microsoft Kinect and Intel RealSense have been popular and employed for various computer graphics applications, including motion capture~\cite{chen2016realtime}, scene reconstruction~\cite{chen20153d}, and image-based rendering~\cite{kraus2007depth}. For such cameras, the resolution and speed of depth acquisition are sacrificed to achieve a low cost. For example, the latest Microsoft Kinect depth camera for Xbox One (Kinect V2) is able to capture depth frames with only 512 $\times$ 424 resolution at 30 frames per second (FPS). While such specifications might be sufficient for certain applications, they are not sufficient for applications involving fast motions and higher frame-rate video. On the other hand, with recent advancements in imaging sensors, high-resolution, high-frame-rate and low-cost video cameras such as GoPro have also opened up many possibilities in computer graphics, such as outdoor motion capture~\cite{shiratori2011motion}, structure from motion (SfM) and dynamic hair capture~\cite{xu2014dynamic}.

Video cameras have their advantages over depth cameras in terms of frame rate and resolution. Observing that high-resolution video cameras are cheap and available anywhere, several techniques (e.g.,~\cite{Park11iccv,Richardt2012CGF}) have been proposed to use a hybrid camera, i.e., a high-resolution video camera and a low-resolution depth camera, to perform spatial upsampling of depth maps. Many applications, such as image-based rendering~\cite{lee2015high,bhat2007using} and image processing~\cite{moreno2007active}, can benefit from additional depth information. The Kinect V2 itself is already such a hybrid camera. However, the low-frame-rate capture problem of existing depth cameras is largely unexplored and thus is the focus of our work.

\begin{figure*}[!t]
\centering
\includegraphics[width=0.9\linewidth]{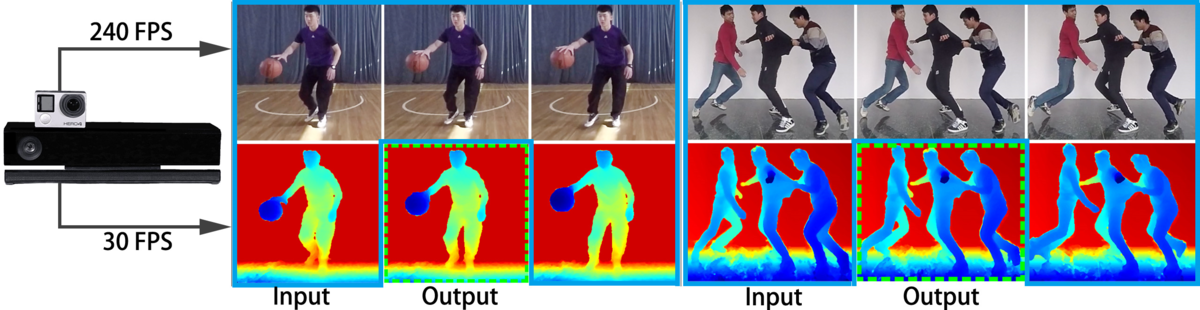}
\caption{Our technique obtains the input via a hybrid camera and is able to temporally upsample the depth maps using a low-frame-rate depth camera with the help of the color images taken by a high-frame-rate video camera. Images surrounded by a cyan line and green dotted line represent the input and output of our method, respectively.}
\label{fig:overview}
\end{figure*}

Motivated by the existing hybrid cameras for obtaining the spatial super-resolution of depth maps and the available high-frame-rate, low-cost video cameras, such as GoPro (with 240 FPS), we propose a hybrid camera to achieve temporal upsampling of depth maps (Fig.~\ref{fig:overview}). Our hybrid camera consists of a low-frame-rate depth camera and a synchronized high-frame-rate video camera. The key challenge is to effectively extract fast motion information from color images using a high-frame-rate video camera and then use it to guide the interpolation of depth maps. A straightforward solution is to first compute the 2D optical flow~\cite{sun2015cvpr} between consecutive images using the high-frame-rate camera and then employ the resulting motion flow to estimate intermediate depth maps between a pair of original depth maps. However, this simple solution works well only for translational motions.

Another possible solution is based on scene flow~\cite{Vedula1999}. However, the traditional methods for scene flow estimation require both color images and depth maps acquired at roughly the same frame rate and thus cannot be directly used for temporal upsampling. To address this problem, we formulate an optimization to estimate the scene flow and intermediate depth maps jointly; the estimated scene flow is used to guide the interpolation of intermediate depth maps, which in turn help refine the scene flow estimation. We derive data constraints from the high-frame-rate color images and enforce spatiotemporal regularization based on the shortest motion path and the locally rigid deformation assumption.

We test our hybrid camera on various examples with quickly moving single or multiple objects and humans. In these challenging examples, which possess non-rigid motions and topology changes, our method has clear advantages over a tracking-based method and the state of the art~\cite{Dolson10cvpr}.
We show that our joint optimization framework can be reduced for scene flow estimation. Compared to the state-of-the-art scene flow methods~\cite{jaimez2015icra,sun2015cvpr,QuirogaBDC14eccv}, our method achieves comparable or even better performance on the MPI Sintel dataset~\cite{Butler:ECCV:2012} and Middlebury stereo dataset~\cite{scharstein2003high}.

\section{Related Work}\label{sec:relatedwork}
Depth cameras are often used together with video cameras to capture RGB-D images. Therefore, the idea of a hybrid camera is not new for 3D imaging. In fact, consumer-level depth cameras such as Kinect V2 are essentially hybrid cameras. It is well known that the depth maps produced by low-cost depth cameras are often noisy and of low resolution. A common approach to enhancing depth maps in the spatial domain is to couple a low-resolution depth map with a high-resolution color image. Various solutions based on optimization (e.g.,~\cite{DiebelT05nips,Park11iccv}), joint edge-preserving upsampling filters (e.g.,~\cite{Yang07cvpr,Theobalt2008ECCV,SongSKH14}), spatiotemporal filtering (e.g.,~\cite{Dolson10cvpr,Richardt2012CGF}), or shading cues (e.g.,~\cite{Wu2014tog,Zollhofer2015tog}) have been explored to increase the spatial resolution of depth maps. All the above methods assume color images and depth maps with the same frame rates. The exceptional case is the work by Dolson et al.~\cite{Dolson10cvpr}. Our experimental results show that our method is more accurate than \cite{Dolson10cvpr} in reconstructing intermediate depth frames. Their method estimates the depth of each pixel by using the time, space, and color information but ignores the depth relationship between adjacent pixels. In contrast, our method not only considers the time, space, and color information but also regularizes the relationship between the depth values of adjacent pixels via the locally rigid priori to approximate the relationship of input depth maps as closely as possible. Further discussions are given in Sec.~\ref{sec:quantitative_evaluation} and ~\ref{sec:qualitative_evaluation}.

Hybrid cameras have also been used for motion deblurring~\cite{Ben-Ezra2004,Li2008,Tai2010pami}. For this application, at least one high-speed but often low-resolution video camera is needed to remove motion blur in color images taken by a low-speed, high-resolution camera. Li et al.~\cite{Li2008} used two low-resolution, high-speed cameras as a stereo pair to reconstruct a low-resolution depth map, the spatial resolution of which was then enhanced by using joint bilateral filters. In addition to reducing motion blur, the approach of Tai et al.~\cite{Tai2010pami} is also used to estimate new high-resolution color images at a higher frame rate. This task of temporal upsampling is similar to ours, but temporal upsampling of depth maps is generally more difficult. Furthermore, the concurrent work of Wang et al.~\cite{wang2017light} used a hybrid camera system consisting of a 3 FPS light field camera and a 30 FPS video camera to reconstruct 30 FPS light field images. The difference between the work of Wang et al. and our method is that they used a learning-based method and upsampled light field images.

While consumer-level depth cameras are able to capture depth maps at only a limited frame rate, high-speed depth cameras, which already reach hundreds or even thousands of frames per second, have been explored in the fields of computer vision and optical engineering in recent years. Among various solutions, structured light illumination (e.g.,~\cite{zhang2006high,narasimhan2008temporal,liu2010dual,gong2010ultrafast,ekstrand2013high,sagawa2014dense}) is the most popular technique, which requires a DLP video projector and a synchronized video camera to acquire structured patterns (e.g., fringe images) projected by a special illuminator. Compared with these approaches, our solution can be regarded as a post-processing technique and is thus applicable to different types of depth cameras. St{\"u}hmer et al.~\cite{jan2015iccv} proposed modifying a typical Time-of-Flight (ToF) camera (e.g., Kinect v2) for model-based tracking at a high frame rate (300 Hz). However, their solution is limited to tracking objects with rigid motion. Our work closely resembles that of Kim and Kim~\cite{kim2014motion}, who used multiview hybrid cameras (consisting of eight high-frame-rate video cameras and six ToF cameras) for motion capture. However, their technique is highly dependent on skeleton tracking and thus is suitable only for articulated motion.

Our joint optimization, which is performed to fuse the color and depth information and estimate the motion field, yields a novel scene flow method. Scene flow estimation for depth cameras is a recent active research topic. For example, Herbst et al.~\cite{herbst2013icra} extended the Horn-Schunck method~\cite{Horn1981} to depth cameras with the depth data term for estimating the scene flow from a consumer-level depth camera. Jaimez et al.~\cite{jaimez2015icra} proposed a total variation regularization term for RGB-D flow estimation in real time. {Piecewise} rigid motion priors were added to the scene flow estimation in~\cite{QuirogaBDC14eccv}. Jaimez et al.~\cite{jaimez20153dv} estimated the scene flow with the joint optimization of motion and segmentation. Their method segments a scene with rigid motions. Sun et al.~\cite{sun2015cvpr} ordered each depth map into layers and assumed the motion {field} in a single layer to be within the small range around the mean rigid rotation. When objects in the same depth layer have large and different motions, this method will introduce artifacts (see Sec.~\ref{sec:quantitative_evaluation}).

As shown in~\cite{QuirogaBDC14eccv,jaimez20153dv}, the piecewise rigid regularization term of the motion field enhances the precision compared with methods such as~\cite{herbst2013icra,jaimez2015icra}. We follow the as-rigid-as-possible energy to model isometric deformations, which have been demonstrated for various graphics applications, such as shape interpolation~\cite{Alexa00}, shape deformation~\cite{ARAP_modeling2007,gao2016efficient,chen2017rigidity} and 3D shape tracking~\cite{zollhoefer2014deformable}. In our work, the as-rigid-as-possible energy is employed for the first time for scene flow estimation with the assumption of nearly isometric deformations.

\section{Hardware Setup}\label{sec:hardware}
Our hybrid camera system is composed of two consumer-level cameras, namely, a GoPro HERO 4 video camera and a Kinect V2 RGB-D camera. The GoPro camera captures color images of WVGA resolution at 240 FPS, while Kinect V2 is able to capture depth maps of $512 \times 424$ resolution at 30 FPS. As shown in Fig.~\ref{fig:overview}, the GoPro is placed above the Kinect, such that Kinect's depth camera is vertically aligned with the GoPro.

\textbf{Calibration and alignment.}\label{sec:align}
The intrinsic and extrinsic parameters of the GoPro and Kinect V2's depth camera are calibrated by the method of~\cite{Zhang2000pami}. The lens distortion of the color images from GoPro is corrected according to the intrinsic parameters of GoPro. We transform the depth maps from the depth camera plane to the color camera plane by using the method introduced by Park et al.~\cite{Park11iccv}. After aligning the depth maps with the undistorted color images, we crop the images to retain the part that needs reconstruction. The two cameras are synchronized in the temporal domain by a flashlight. More specifically, we capture the flashlights and then identify the first highlighted images from the color and depth sequences that refer to the same point in time. Finally, we acquire the aligned and synchronized depth maps, color images, and camera intrinsic parameters of the cropped and aligned images, which are denoted by $\mathbf{D}$, $\mathbf{C}$ and $\mathbf{A}$, respectively.

\section{Temporal Upsampling of Depth maps}\label{sec:recons}

\begin{figure}[!t]
\centering
\subfloat[$\mathbf{C}_{t(k)}$]{\includegraphics[width=0.18\linewidth]{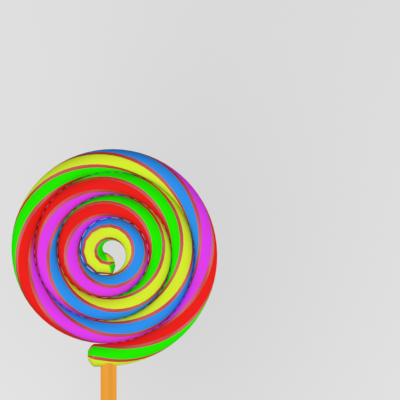}}\hspace{0.1ex}
\subfloat[$\mathbf{C}_{t(k)+1}$]{\includegraphics[width=0.18\linewidth]{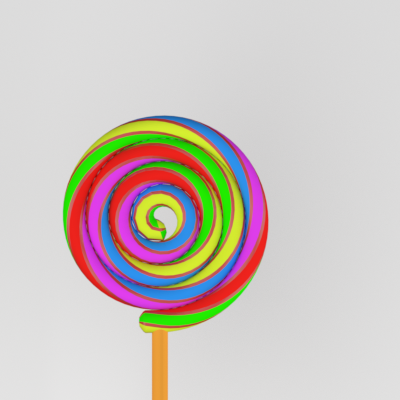}}\hspace{0.1ex}
\subfloat[$\mathbf{C}_{t(k)+2}$]{\includegraphics[width=0.18\linewidth]{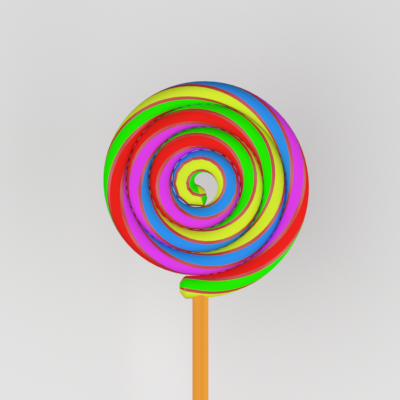}}\hspace{0.1ex}
\subfloat[$\mathbf{C}_{t(k)+3}$]{\includegraphics[width=0.18\linewidth]{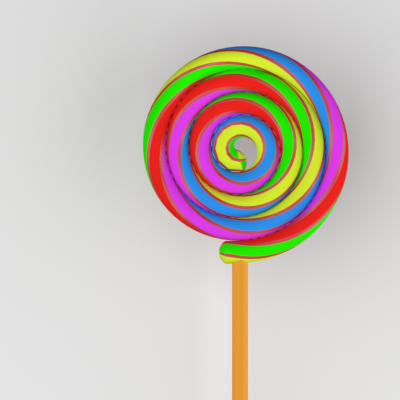}}\hspace{0.1ex}
\subfloat[$\mathbf{C}_{t(k)+4}$]{\includegraphics[width=0.18\linewidth]{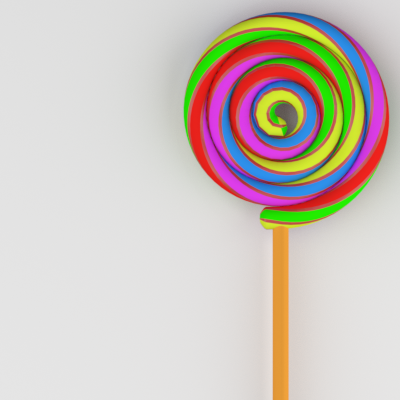}}
\\
\subfloat[$\mathbf{D}_k$]{\includegraphics[width=0.18\linewidth]{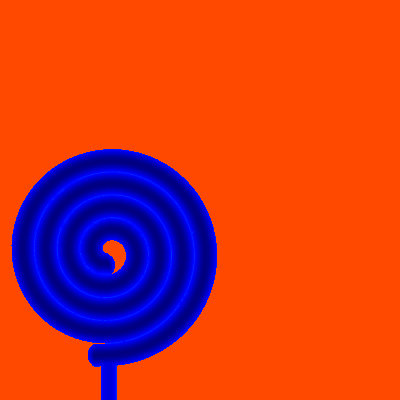}}\hspace{0.1ex}
\subfloat[$\mathbf{D}_{t(k)+1}$]{\includegraphics[width=0.18\linewidth]{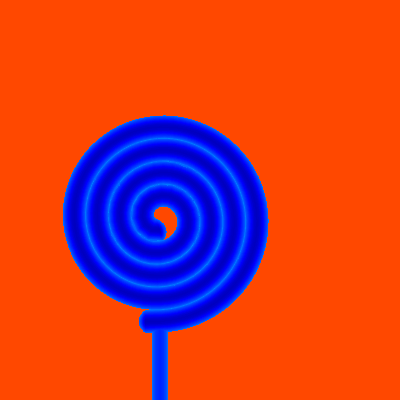}}\hspace{0.1ex}
\subfloat[$\mathbf{D}_{t(k)+2}$]{\includegraphics[width=0.18\linewidth]{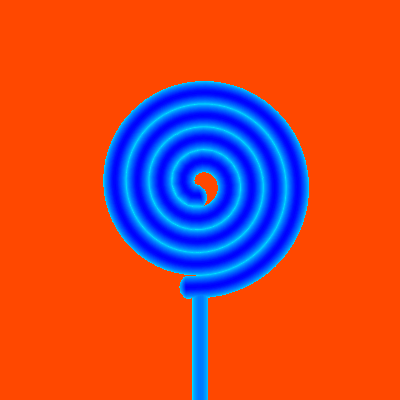}}\hspace{0.1ex}
\subfloat[$\mathbf{D}_{t(k)+3}$]{\includegraphics[width=0.18\linewidth]{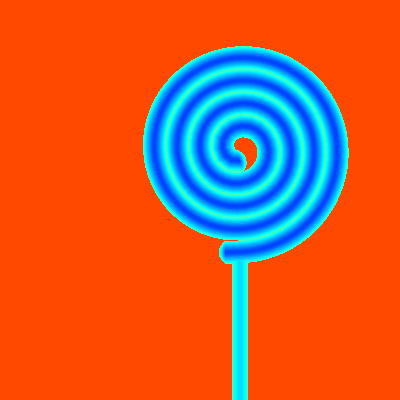}}\hspace{0.1ex}
\subfloat[$\mathbf{D}_{k+1}$]{\includegraphics[width=0.18\linewidth]{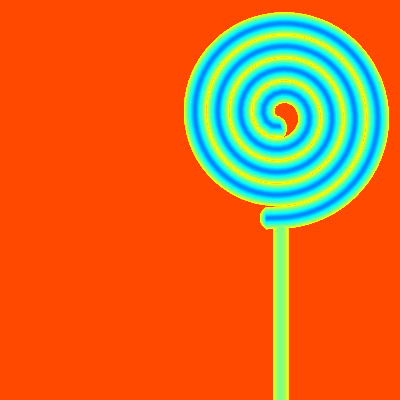}}
\caption{Our main idea illustrated on synthetic chronological data. Given input color images (a-e) and depth maps (f) and (j), we temporally upsample depth frames by reconstructing intermediate depth maps (g-i).}
\label{fig:synthinput}
\end{figure}

\begin{figure}[!t]
\centering
\includegraphics[width=0.90\linewidth]{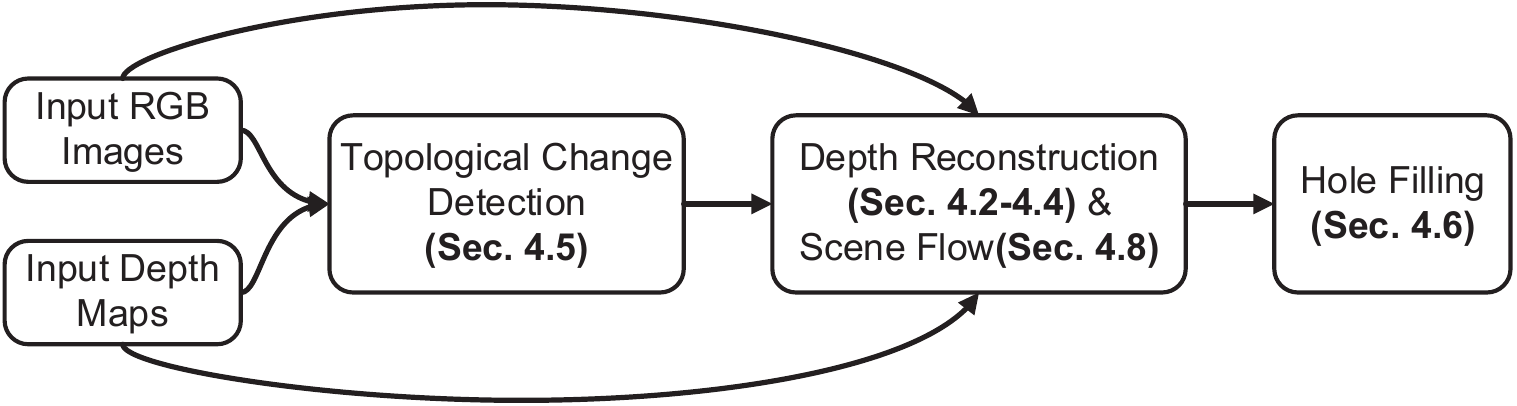}
\caption{Pipeline of our system.}
\label{fig:pipeline}
\end{figure}

Our main goal is to temporally upsample the depth maps by estimating a depth map corresponding to each color image from a higher-frame-rate video camera, as illustrated in Fig.~\ref{fig:synthinput}. The optical flow term is employed to exploit the dense 2D motion information from the color images. To connect between the 3D motion and 2D optical flow, a projection term is employed. We use the popular point-to-point and point-to-plane terms to exploit the start and end positions of the depth maps reconstructed from the two consecutive depth maps captured using Kinect V2 (Sec.~\ref{sec:optical_flow}). We also employ regularization terms to enforce the local rigidity and shortest path in the motion flow (Sec.~\ref{sec:spatialregu}). Considering occlusion in the motion, we apply occlusion detection to avoid artifacts resulting from the occluded regions (Sec.~\ref{sec:occdetect}). The precalculated optical flow is used to detect the topology changes (Sec.~\ref{sec:topochange}). Moreover, to fill the remaining holes, we use both forward and backward reconstruction and a bilateral filter (Sec.~\ref{sec:holefilled}). We use a joint optimization to determine the above data constraints and spatiotemporal regularization terms (Sec.~\ref{sec:energy_minimization}). Finally, the framework is reduced to a scene flow method with pairs of color images and depth maps (Sec.~\ref{sec:sceneflowestim}). The pipeline of our system is shown in Fig.~\ref{fig:pipeline}. All the above components work together to reconstruct the depth maps and estimate the scene flow. The importance of each component will be evaluated in Sec.~\ref{sec:impeachterm}.

\subsection{Notations}\label{sec:enegy}

Given a pair of consecutive depth frames $\mathbf{D}_{k}$ and $\mathbf{D}_{k+1}$, as illustrated in Fig.~\ref{fig:synthinput}, let $\mathbf{C}_{t(k)}$ denote the color image corresponding to $\mathbf{D}_{k}$, $\mathbf{C}_{t(k) + g}$ denote the color image corresponding to $\mathbf{D}_{k+1}$, and $\mathbf{C}_{t(k) + s}$ ($s{=}1, \cdots,g{-}1$) denote the intermediate color images captured between $\mathbf{C}_{t(k)}$ and $\mathbf{C}_{t(k) + g}$. $g{=}4$ for the synthetic example in Fig.~\ref{fig:synthinput}. Our ultimate goal is thus to reconstruct a depth map $\mathbf{D}_{t(k) + s}$ corresponding to $\mathbf{C}_{t(k) + s}, s \in [1,g{-}1]$. Our underlying optimization will also reconstruct $\mathbf{D}_{t(k) + g}$ to exploit the boundary constraints from $\mathbf{D}_{k+1}$ in depth reconstruction. $\mathbf{D}_{t(k) + g}$ will be replaced with $\mathbf{D}_{k+1}$, the depth information of which is accurate. The point cloud $\mathbf{M}_{k}$ is generated by projecting $\mathbf{D}_k$ to the 3D space with the intrinsic parameters $\mathbf{A}$. The above and additional notations are summarized in Tab.~\ref{tab:notations}. In our setting of the capture rate ($240$ FPS) for the GoPro and that ($30$ FPS) for Kinect V2, the value of $g$ is $8$.

\renewcommand{\multirowsetup}{\centering}
\begin{table}[!h]
\caption{Notations.}
\label{tab:notations}
\centering
\begin{tabular}{c|p{0.7\linewidth}}
\cline{1-2}
$\textbf{Notation}$ & $\textbf{Exposition}$ \\
\cline{1-2}
$k$ & the index of the captured depth maps \\
\cline{1-2}
\multirow{2}{2cm}{$g$}  & the ratio of the color camera's FPS to the depth camera's FPS \\
\cline{1-2}
\multirow{2}{2cm}{$\mathbf{D}_k$, $\mathbf{D}_{k+1}$} & the first and second of a pair of consecutive depth maps generated by Kinect V2\\
\cline{1-2}
\multirow{3}{2cm}{$\mathbf{D}_{t(k)+s}$} &  a depth map to be reconstructed corresponding to $\mathbf{C}_{t(k)+s}$, $1 \leq s \leq g$\\
\cline{1-2}
\multirow{3}{2cm}{$\mathbf{C}_{t(k)+s}$} &  color images captured at the interval between $\mathbf{D}_k$ and $\mathbf{D}_{k+1}$, $0 \leq s \leq g$, with $\mathbf{C}_{t(k)}$ corresponding to $\mathbf{D}_k$ and $\mathbf{C}_{t(k)+g}$ to $\mathbf{D}_{k+1}$\\
\cline{1-2}
\multirow{2}{2cm}{$\mathbf{p}_{s,i}$} & the position of the $i$-th point in the point cloud corresponding to $\mathbf{D}_{t(k)+s}$; $\mathbf{P} = \{\mathbf{p}_{s, i}\}$\\
\cline{1-2}
\multirow{3}{2cm}{$\mathbf{R}_{s,i}$} & a rotation matrix associated with the $i$-th point in the point cloud corresponding to $\mathbf{D}_{t(k)+s}$; $\mathbf{R} = \{\mathbf{R}_{s, i}\}$\\
\cline{1-2}
\multirow{2}{2cm}{$\mathbf{v}_{s,i}$} & optical flow at the $i$-th pixel in $\mathbf{C}_{t(k)+s}$, $0 \leq s \leq g$; $\mathbf{v}_s$ is the optical flow for $\mathbf{C}_{t(k)+s}$\\
\cline{1-2}
$\mathbf{M}_k$,$\mathbf{M}_{k+1}$ & point cloud generated by $\mathbf{D}_k$ and $\mathbf{D}_{k+1}$\\
\cline{1-2}
\end{tabular}
\end{table}

\subsection{Data Terms}\label{sec:optical_flow}
To exploit the motion information from the color images, we employ the optical flow data term, point-to-point term, point-to-plane term and projection term as the motion estimation constraints. To estimate the optical flow at the interval between the consecutive depth maps $\mathbf{D}_{k}$ and $\mathbf{D}_{k+1}$, we use the color images $\mathbf{C}_{t(k)+s}$, $s \in [0,g]$, to recover the 2D motion flow between $\mathbf{D}_k$ and $\mathbf{D}_{k+1}$. The optical flow data term is shown below:
\begin{equation}\label{eqn:opti}
\begin{split}
& \quad E_{opti}(\mathbf{v}_s) \\
& =\lambda_{opti}\sum_{\mathbf{s}\in [0,g-1]}\sum_{\mathbf{x},\mathbf{y}}\rho( \mathbf{C}_{t(k)+s}(\mathbf{x}+\mathbf{v}_{\mathbf{s},\mathbf{x}},\mathbf{y}+\mathbf{v}_{\mathbf{s},\mathbf{y}}) \\
& - \mathbf{C}_{t(k)+s +1}(\mathbf{x},\mathbf{y})),
\end{split}
\end{equation}
where $\mathbf{v}_{\mathbf{s},\mathbf{x}}$ and $\mathbf{v}_{\mathbf{s},\mathbf{y}}$ are the optical flow in the images along the $x$-axis and $y$-axis, respectively. $\lambda_{opti}$ is the weight of the optical flow term and is chosen to be 8 in the implementation. $\rho(r){=}\sqrt{r^{2}+\varepsilon^{2}}$ $\label{eqn:robustfunction}$ is the kernel function that defines the robust metric for addressing noise and outliers ($\varepsilon{=}10^{-4}$ in the implementation)~\cite{black1993a,brox2004high}. Furthermore, to improve the optical flow value, we employ the weighted median filter to avoid over-smoothing along object edges~\cite{sun2010cvpr}.

The previous RGB-D scene flow methods~\cite{sun2015cvpr,QuirogaBDC14eccv,jaimez2015icra} use local or global smooth terms to solve the undetermined problem in the depth map field. In this work, we lift the depth similarity constraint to the 3D space where the geometry information can be explored better~\cite{chen20153d}. The advantage of such a distance metric in the 3D Euclidean space instead of in the depth difference is that not only the local geometric distance but also the surface normal information can be employed to measure the geometric distance. We project depth map $\mathbf{D}_k$ to the 3D space to generate a point cloud $\mathbf{M}_K$. The $i$-th pixel in $\mathbf{D}_k$ is projected to $\mathbf{p}_i$ in 3D. The connecting relationship between pixel $i$ and its adjacent pixels is created for the spatial coherency.

We employ the following point-to-point term and point-to-plane term~\cite{pottmann2006geometry} to reconstruct the geometry constraints of the depth map $\mathbf{D}_{t(k)+g}$:
\begin{equation}\label{eqn:point}
E_{point}({\mathbf{p}_{g,i}}) =\lambda_{point}\sum_{i \in V}{\|{\mathbf{p}_{g,i}}-\tilde{\mathbf{p}}_{k+1,i}\|_2^2},
\end{equation}
\begin{equation}\label{eqn:plane}
E_{plane}({\mathbf{p}_{g,i}}) =\lambda_{plane}\sum_{i \in V}{\|\mathbf{n}_{k+1,i}^{T}({\mathbf{p}_{g,i}}-\tilde{\mathbf{p}}_{k+1,i})\|_2^2},
\end{equation}
where $\tilde{\mathbf{p}}_{k+1,i}$ is the closest point to $\mathbf{p}_{k,i}$ in $\mathbf{M}_{k+1}$ and $\mathbf{n}_{k+1,i}$ is the normal vector of $\tilde{\mathbf{p}}_{k+1,i}$. The energy weights for the point-to-point term $\lambda_{point}$ and point-to-plane term $\lambda_{plane}$ are set to $9$ and $10$, respectively. Since the optical flow is essentially the projection of the scene flow introduced by the reconstructed depth maps, the projection function $\psi(\cdot)$ can project the point to the video camera's plane by $\mathbf{A}$~\cite{Park11iccv}. We model this constraint as follows:
\begin{equation}\label{eqn:project}
\begin{split}
& \quad E_{proj}(\mathbf{p}_{s,i}, \mathbf{v}_{s,i})
\\
&=\lambda_{proj}\sum_{s \in {[1, g-1]}}\sum_{i \in V}\mathbf{O}({\mathbf{v}_{s,i}},\mathbf{C}_{t(k)+s}){\|\psi({\mathbf{p}_{s+1,i}} -{\mathbf{p}}_{s,i})-\mathbf{v}_{s,i}\|_2^2}
\end{split}
\end{equation}
and use the projection term to connect the optical flow and 3D point cloud. The energy weight $\lambda_{proj}$ is set to $5$. $\mathbf{O}(\cdot)$ is a function that indicates whether this point is occluded in motion (Sec.~\ref{sec:occdetect}).

\subsection{Spatial and Temporal Regularization}\label{sec:spatialregu}
With the observation that most of the objects in real-world scenes move in a rigid or locally rigid fashion, we employ the following locally rigid regularization term:

\begin{equation} \label{eqn:regularity}
\begin{split}
& \quad {E_{rigid}(\mathbf{p}_{s,i}, \mathbf{R}_{s,i}) }\\
& =\lambda_{rigid}\sum_{s \in {[1,g]}}\sum_{i \in V} {\sum_{j \in N_i}w_{ij}{{ \| ({\mathbf{p}_{s,i}}\,\,-{\mathbf{p}_{s,j}}) - \mathbf{R}_{s,i}\, (\mathbf{p}_{k,i}-\mathbf{p}_{k,j}) \|}^2_2}},
\end{split}
\end{equation}
where $N_i$ denotes points connected to the $i$-th point in the point cloud and $\lambda_{rigid}$ is set to $16$. It is more likely that the connected points with similar depth and color would share similar locally rigid motions. These weights $w_{ij}$ are defined as $w_{c,ij} \cdot w_{d,ij} \cdot w_{t,ij}$, with the depth coherence $w_{d,ij}=\exp(-{{\|\mathbf{p}_{k,i}-\mathbf{p}_{k,j}\|}^{2}/{\sigma_d^{2}}})$, color coherence $w_{c,ij}=\exp(-{{\|\mathbf{C}_{t(k),i}-\mathbf{C}_{t(k),j}\|}^{2}/{\sigma_c^{2}}})$, and topology change $w_{t,ij}=\exp(-{{\|\mathbf{e}_{i',j'}-\mathbf{e}_{i,j}\|}^2}/{\sigma_t^2})$, where $\mathbf{e}_{i,j}$ and $\mathbf{e}_{i',j'}$ are the Euclidean distance of the corresponding point pair $(i,j)$ in $\mathbf{M}_{k}$ and warped $\mathbf{M}_{k}$, respectively (see Tab.~\ref{tab:notations} and Sec.~\ref{sec:topochange}). The connection of occlusive points is generally less reliable. When at least one of $\mathbf{p}_i$ and $\mathbf{p}_j$ is occluded or out of the image boundary and thus does not have the corresponding point pair in $\mathbf{M}_{k+1}$, $w_{t,ij}$ is set to 1. In practice, $\sigma_t = 0.015$, $\sigma_c = 1$ and $\sigma_d = 0.015$. The total quadratic variations are employed to regularize the motion field and prevent an artifact being generated in large-scale deformation~\cite{levi2015smooth}. This is defined as follows:
\begin{equation}\label{eqn:smooth}
E_{reg}(\mathbf{R}_{s,i}) =\lambda_{reg}\sum_{s \in [1,g]}\sum_{i \in V} {\sum_{j \in N_i}w_{ij}{\|\mathbf{R}_{s,i}-\mathbf{R}_{s,j}\|}^2_2},
\end{equation}
where $\lambda_{reg}$ is set to 0.8.

The temporal regularization term is employed to reduce the uncertainty of the solution and to favor the solution with the shortest path. This term is defined as the sum of the Euclidean distances of the corresponding points in the consecutive point cloud:
\begin{equation}
E_{short}(\mathbf{p}_{s,i}) = \lambda_{short}\sum_{s \in [1,g]}\sum_{i \in V}\|\mathbf{p}_{s,i}-\mathbf{p}_{s-1,i}\|^2_2,
\end{equation}
where $\lambda_{short}$ is set to $7$.

\subsection{Occlusion Detection}\label{sec:occdetect}
Due to different objects or different parts of the same object moving with different speeds, there often exist occlusions in the camera view. The occlusion leads to three problems.

\begin{figure}[!t]
\centering
\subfloat[]{\includegraphics[width=0.21\linewidth]{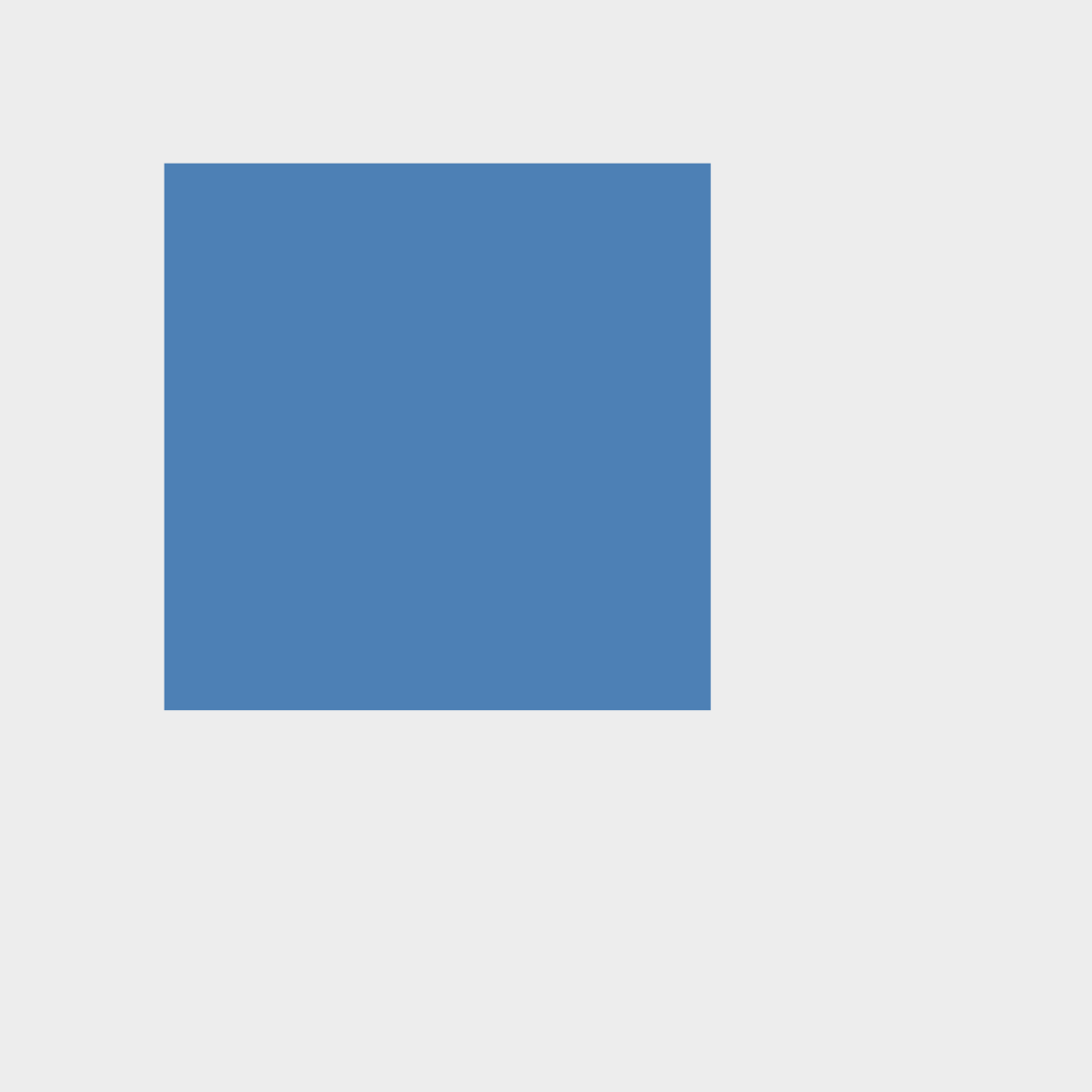}}\hspace{2em}
\subfloat[]{\includegraphics[width=0.21\linewidth]{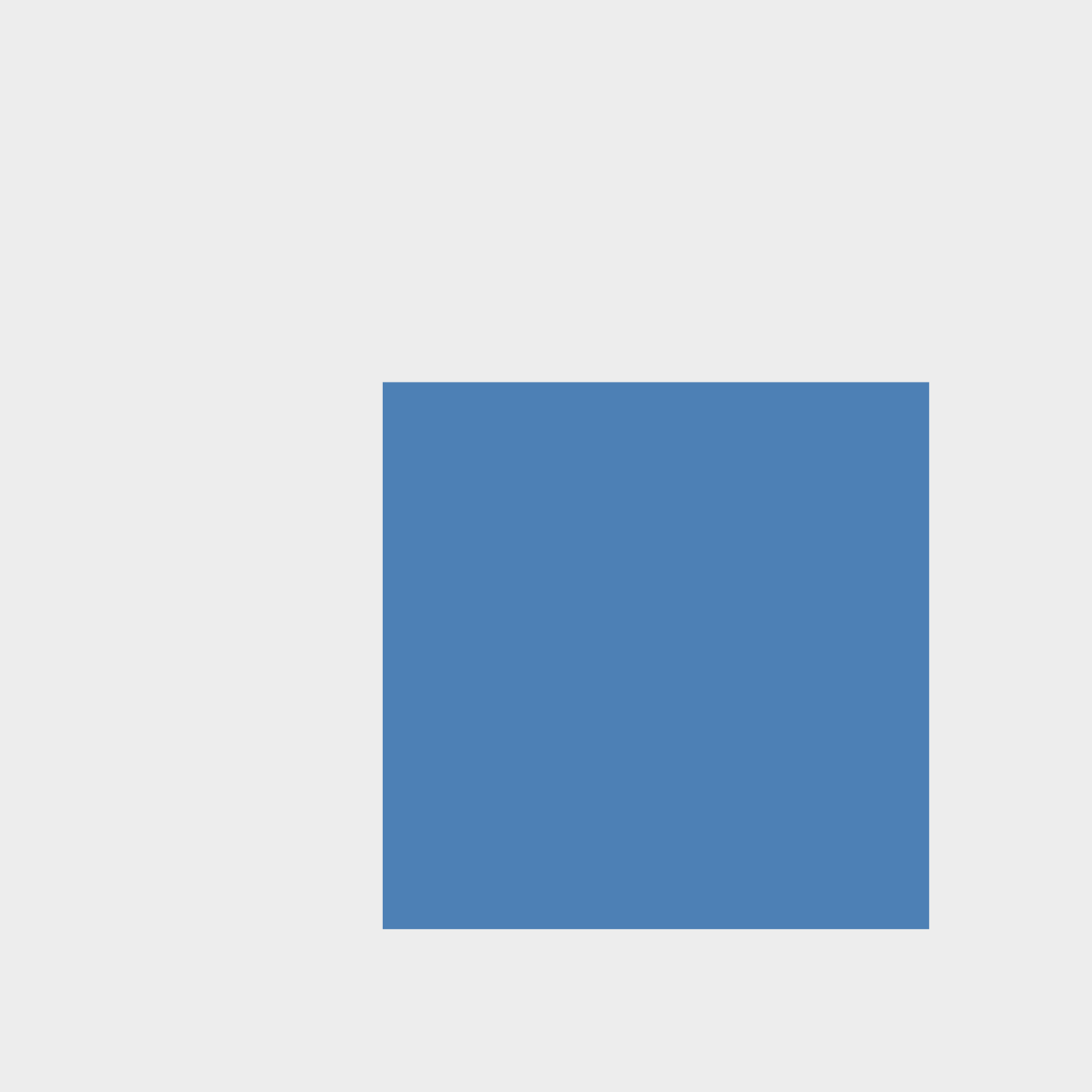}}\hspace{2em}
\subfloat[]{\includegraphics[width=0.21\linewidth]{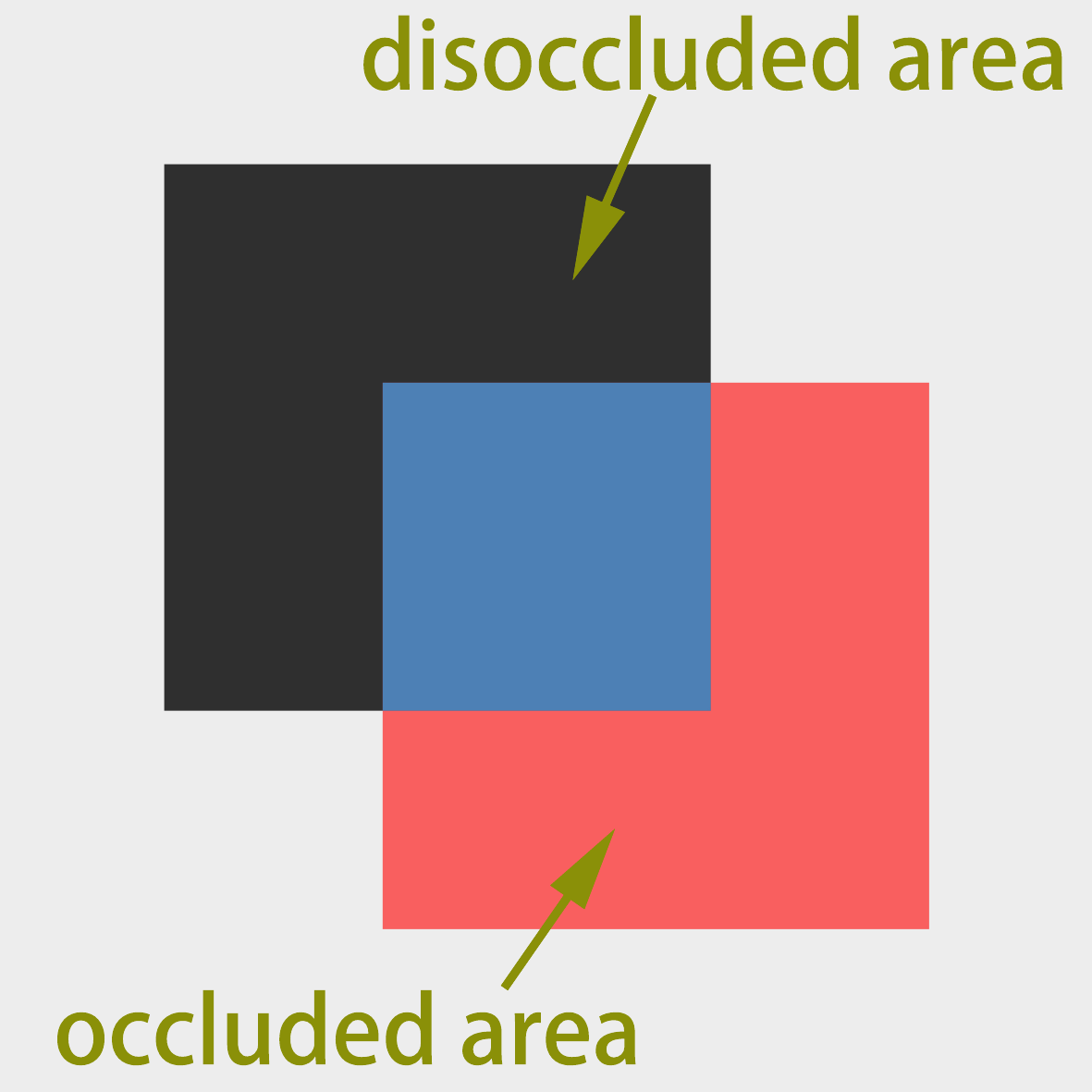}}
\caption{The blue block in (a) moves to the lower right, resulting in (b). (a) is warped by the optical flow; thus, (c) is generated. The black region in (c) is the disoccluded area of (a), while the red region is the occluded area. The red region in (c) appears in (a) but is covered in (b), while the black area in (c) is present in (b) but is covered in (a).}
\label{fig:twokinesocclusion}
\end{figure}

First, the occlusion results in the projection relationship error in $E_{proj}$, as the occlusion degrades the relation between the 3D motion flow and the optical flow. To reduce the impact of the mismatch of the projection relation, we adopt the idea in~\cite{sand2006particle} to detect the occlusion based on flow divergence and the pixel difference. The $\mathbf{O}(\mathbf{v}_{s,i},\mathbf{C}_{t(k)+s})$ function used in Eq.~\ref{eqn:project} is defined as $\exp(-{{\|\mathbf{div}(s,i)\|}^2}/{\sigma_1^2}) \cdot \exp(-{{\|\mathbf{C}_{t(k)+s,i}-\mathbf{C}_{t(k)+s+1,i+\mathbf{v}_{s,i}}\|}^2}/{\sigma_2^2})$, where $i+\mathbf{v}_{s,i}$ is the index of the pixel obtained by applying the translation of $\mathbf{v}_{s,i}$ to the index $i$ ~\cite{barivcevic2017user}. $\mathbf{div}(s,i)$ is the divergence of the optical flow. Based on our experience, we set $\sigma_1 = 1$ and $\sigma_2 = 20$. The motion of its connected points will yield the occluded depth pixels without the need for optical flow information.

Second, the occlusion also produces an outlier of $E_{opti}$ due to color pixel mismatch~\cite{xiao2006bilateral}. To address the outlier, we use the robust kernel function $\rho(r)$ in Sec.~\ref{sec:optical_flow}. However, the outlier still makes the optical flow over-smoothed at the boundaries~\cite{sun2010cvpr}. This problem can be further alleviated by applying the weighted media filter~\cite{sun2010cvpr}.

Third, the occlusion generates holes in the reconstructed depth maps. The occluded surfaces are divided into occlusion and disocclusion areas, as illustrated in Fig.~\ref{fig:twokinesocclusion}~\cite{Black:CVPR:1991}. The disoccluded objects can be detected in the reversed timeline. In other words, the portion of the disoccluded objects can be recovered from the corresponding frames in the backward-reconstructed depth maps sequence. We will show how these holes can be filled in Sec.~\ref{sec:holefilled}.

\subsection{Topology Change Detection}\label{sec:topochange}
The connection between neighboring pixels describes an object's topology. Topology changes often occur in scene objects that interact with each other via a clap, rebound or handshake. In such interactions, the separation of points, or the combination of points, is regarded as a topology change. Both cases will change the connection of points at an object boundary.

\begin{figure}[!t]
\centering
\subfloat[]{\includegraphics[width=0.28\linewidth]{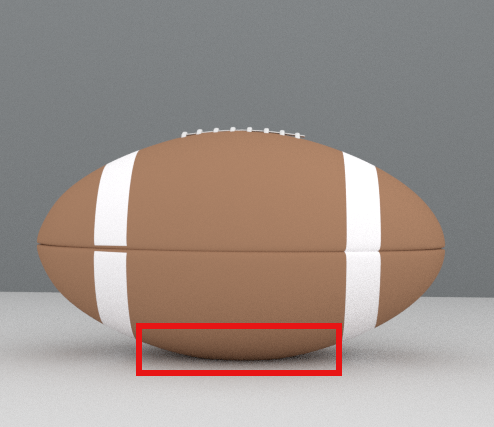}}\hspace{0.5em}
\subfloat[]{\includegraphics[width=0.28\linewidth]{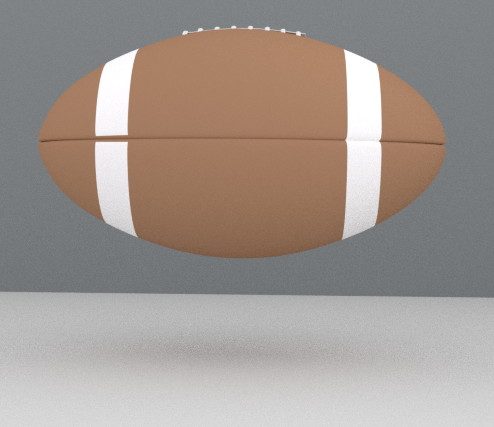}}\hspace{0.5em}
\subfloat[]{\includegraphics[width=0.28\linewidth]{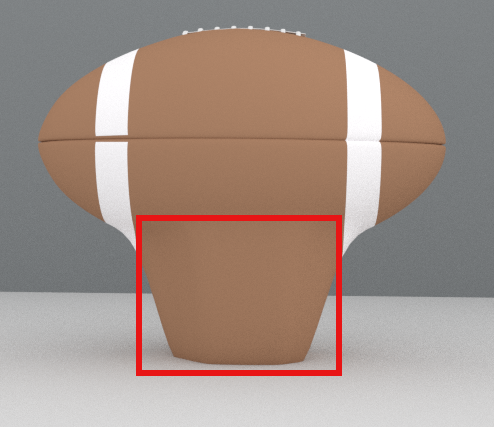}}
\caption{Topology change detection. The football in (b) is the football in (a) after having bounced from the ground. (a) is warped by 3D motion flow to (c). The red insets of (a) and (c) present the areas in which the topology changes.
}
\label{fig:topologdetect}
\end{figure}

The topology changes due to the merging of separated objects are solved by the point cloud stitching, with use of the information regarding the optical flow and geometry constraints (Sec.~\ref{sec:optical_flow}). We detect the topology changes of separating objects by computing the distance change, with respect to the adjacent points, between the point cloud and the warped one. The latter is obtained by warping the original point cloud using the rough motion flow in 3D space from $\mathbf{D}_{k}$ to $\mathbf{D}_{k+1}$, as shown in Fig.~\ref{fig:topologdetect}. To estimate the proximate motion flow, we project the optical flow that is accumulated from ${\mathbf{C}_{t(k)}}$ to ${\mathbf{C}_{t(k)+g-1}}$ to 3D space with $\mathbf{A}$ and $\mathbf{D}_k$. To keep the geometric information of the warped point cloud consistent with that of the original one, we use $E_{point}$, $E_{plane}$, and $E_{rigid}$ to obtain a coarse result. As mentioned in Sec.~\ref{sec:spatialregu}, we define a weight of topology change $w_{t,ij}$ to represent topology changes between a pair of points $(i,j)$.

\subsection{Hole Filling}\label{sec:holefilled}
\begin{figure*}[!t]
\centering
\includegraphics[width=0.98\linewidth]{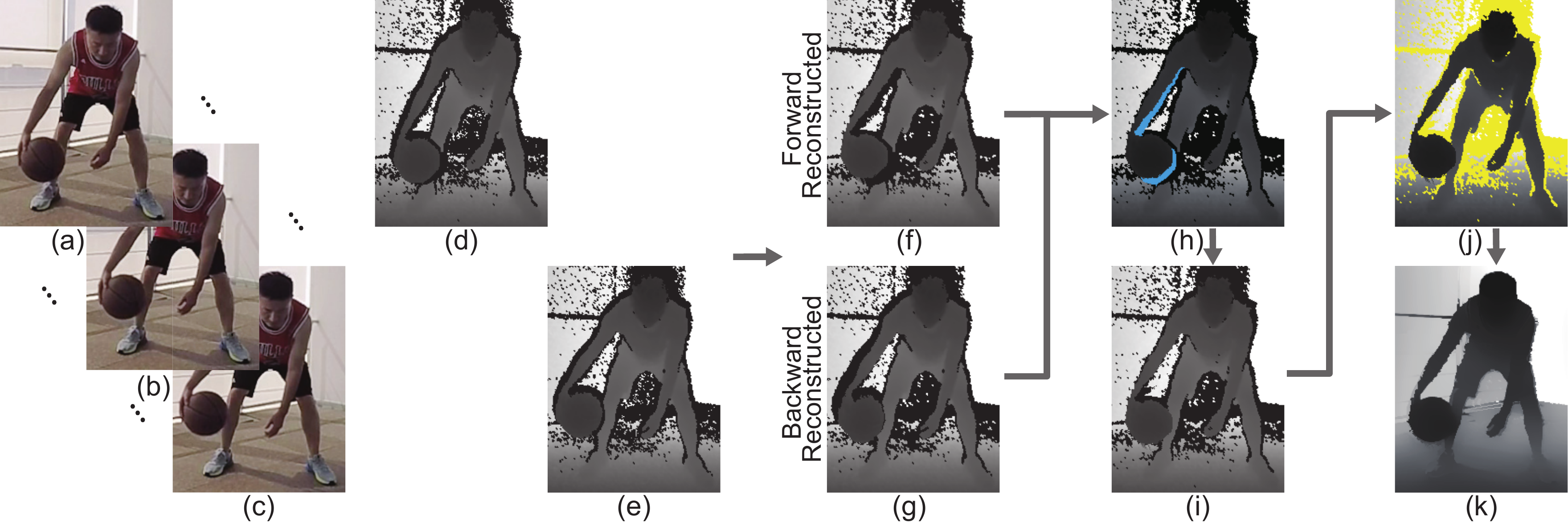}
\caption{Hole filling. (a-c) are the input color images. (d) and (e) are the input depth maps corresponding to (a) and (c), respectively. (f) is the forward-reconstructed depth map corresponding to (b), and (g) is the backward-reconstructed result. Holes exist in the reconstructed depth maps (f) and (g) due to either occlusion or the imaging principles. In (h), the pixels in cyan represent the first type of holes generated by occlusion. (i) is the depth map fixed using forward- and backward-reconstructed depth maps. The yellow pixels of (j) are the second type of holes, which occur due to the imaging artifacts in (i).
(k) is the result after filling all the holes.}
\label{fig:holefill}
\end{figure*}

To reconstruct the depth maps more completely, we fill the holes in the depth maps. Holes occur due to either the occlusion or the imaging principle of a depth camera. In Fig.~\ref{fig:holefill} (h), the cyan pixels exemplify the first type of holes, which are caused by occlusion. The yellow pixels exemplify the second type of holes, which natively exist in the input depth maps. We use different approaches to fill these two kinds of holes. The workflow is shown in Fig.~\ref{fig:holefill}.

To address the first type of holes, we use the forward and backward reconstruction information together~\cite{l0Norigid2015}. During the forward reconstruction, the depth data of the occluded portion in the initial frame are also missing in the next $g{-}1$ forward reconstructed depth maps. The occluded part of the reconstructed depth data in the current frame is thus the accumulated occluded part of the previous depth frames in the current forward reconstructed sequence. On the other hand, the occluded part of the forward reconstruction is the disoccluded part of backward reconstruction. Thus, the missing depth information of forward reconstructed maps can be recovered from the backward reconstruction depth maps. By comparing the final depth frame of forward reconstruction ($\mathbf{D}_{t(k)+g}$) with the initial depth map of backward reconstruction ($\mathbf{D}_{k+1}$), we can obtain the corresponding relationship between the disocclusive depth data and the pixels in the backward reconstructed depth map and thus fill in the holes of this type, as shown in Fig.~\ref{fig:holefill}(i).

The second type of holes appears due to two main reasons: imperfect alignment of the depth and color images and environmental interference, such as hair, glare, motion blur and the reflectivity of objects. Some of the holes of this second type can also be filled using forward- and backward-reconstructed depth maps. This is because the missing depth data of the second type of holes can be obtained from adjacent depth maps by Kinect V2. We use the bilateral filter, with the help of the color information, to fill in the residual missing depth data~\cite{Theobalt2008ECCV}.

\subsection{Energy Minimization}\label{sec:energy_minimization}

At every interval between two consecutive Kinect V2 captured depth maps, we reconstruct the intermediate depth maps by optimizing the following global energy Eq.~\ref{eqn:energy}, which consists of the energy terms introduced in the previous section:

\begin{align} \label{eqn:energy}
\begin{split}
& \quad E({\mathbf{P}},\mathbf{V},\mathbf{R}) \\
& =  E_{opti}(\mathbf{v}_s) + E_{point}({\mathbf{p}_{g,i}}) + E_{plane}({\mathbf{p}_{g,i}}) \\
& \quad + E_{proj}(\mathbf{p}_{s,i}, \mathbf{v}_{s,i}) + E_{rigid}(\mathbf{p}_{s,i}, \mathbf{R}_{s,i})\\
& \quad + E_{reg}(\mathbf{R}_{s,i}) + E_{short}(\mathbf{p}_{s,i}).
\end{split}
\end{align}

This equation can be rewritten as the following summation of squared residues: $\mathbf{E}(x){=}\sum_{i}g_i(\mathbf{x})^{2}{=}\|\mathbf{g}(\mathbf{x})\|^{2}_2,$ where $\mathbf{x}$ is a vector of all variables and $\mathbf{g}(\mathbf{x})$ is a vector function, with its element function denoted by $g_i(\mathbf{x})$. Minimizing $\mathbf{E}(x)$ is a least squares problem. The Gauss-Newton method is employed to solve this optimization~\cite{wu2016simultaneous}. In the $k$-th Gauss-Newton iteration, we update the variables according to $\mathbf{x}_{k+1}{=} \mathbf{x}_{k}{+}\Delta{\mathbf{x}}$. $\Delta{\mathbf{x}}$ is satisfied by the equation $\mathbf{J}^{T}(\mathbf{x}_{k})\mathbf{J}(\mathbf{x}_{k})\Delta{\mathbf{x}}{=}{-}\mathbf{J}^{T}(\mathbf{x}_{k})\mathbf{g}(\mathbf{x}_{k})$, where $\mathbf{J}(\mathbf{x}_{k})$ is the Jacobian matrix at ${\mathbf{x}_{k}}$. To solve these linear equations, we apply the preconditioned conjugate gradient (PCG) solver with several CUDA kernels~\cite{zollhoefer2014deformable}.

The optical flow is initialized by the GPU-based method~\cite{bworld}. In this optimization, we use three hierarchical levels. During the prolongation from the coarse level to the fine level, the bilinear interpolation is applied to the optical flow and point cloud. The rigid rotation in $\mathbf{E}_{rigid}$ remains the same in the corresponding position in the coarse levels.

\subsection{Scene Flow Estimation}\label{sec:sceneflowestim}
This joint optimization framework will introduce the depth maps and the scene flow simultaneously. If our input is a set of color images, each of which has its corresponding depth map, our method reduces to a standard scene flow method. In Sec.~\ref{sec:quantitative_evaluation}, we give quantitative evaluations of our scene flow method and show its advantage over the current state-of-the-art works~\cite{jaimez2015icra,QuirogaBDC14eccv,sun2015cvpr}.

\section{Results and Discussions} \label{sec:results}

In this section, we make both qualitative and quantitative evaluations of our approach. Our technique was tested on real-world complex scenes such as those depicting basketball \& games. There are many challenges in real-world scenes, including occlusion, topology changes and moving cameras. These challenges are demonstrated in different cases in the following sections. From the visual results, it is clear that our method performs better than the state-of-the-art methods. These real-world scenes lack the ground truth of scene flow and depth maps. To perform the quantitative evaluations, the MPI Sintel dataset~\cite{Butler:ECCV:2012} and Middlebury stereo dataset~\cite{scharstein2003high} with ground truth are employed.

\textbf{Performance.}
The performance of our hybrid camera system was measured on an Intel Xeon E5-2520 CPU with 32 GB of RAM and a single NVIDIA Titan X. Between every pair of successive depth frames, we reconstructed the missing depth maps (8 frames in total) according to the color images taken by the GoPro. This whole process took approximately 315.6 seconds on average for all the tested sequences. At the coarse, medium, and fine levels, the duration of an iteration was approximately 2 seconds, 3 seconds, and 15 seconds, respectively. At the coarse level, the total time of more than ten iterations was 30 seconds. This time was 45 seconds for the medium level and 255 seconds for the fine level. All three levels included 5 Gauss-Newton iterations (within each iteration, there were 10 PCG iterations). The average time for reconstructing one depth frame was 39.5 seconds.

\begin{figure*}[!h]
\centering
\subfloat[$\lambda_{rigid}$]{\includegraphics[width=0.173\linewidth]{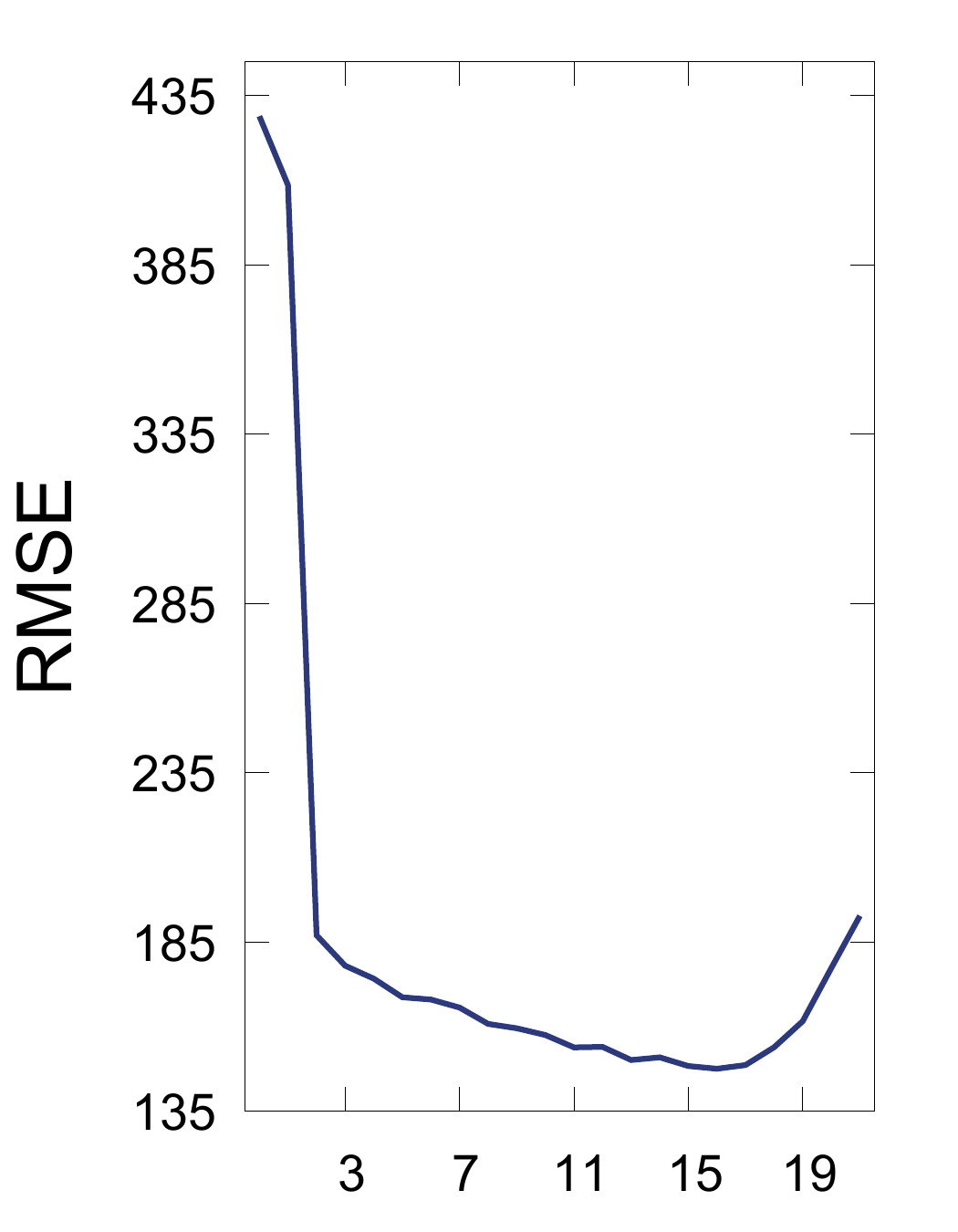}}\hspace{-1.1em}
\subfloat[$\lambda_{opti}$]{\includegraphics[width=0.16\linewidth]{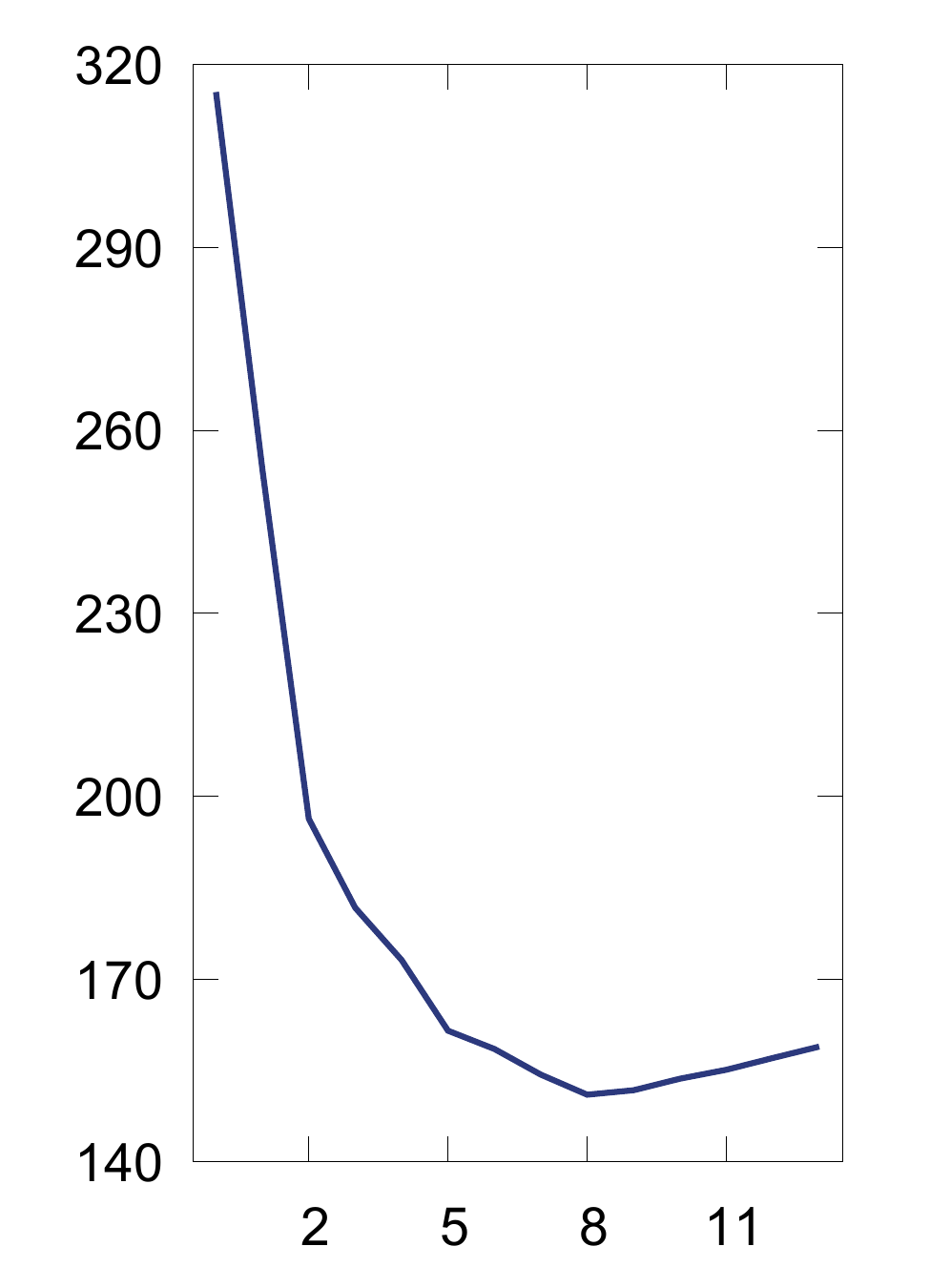}}\hspace{-1.1em}
\subfloat[$\lambda_{plane}$]{\includegraphics[width=0.16\linewidth]{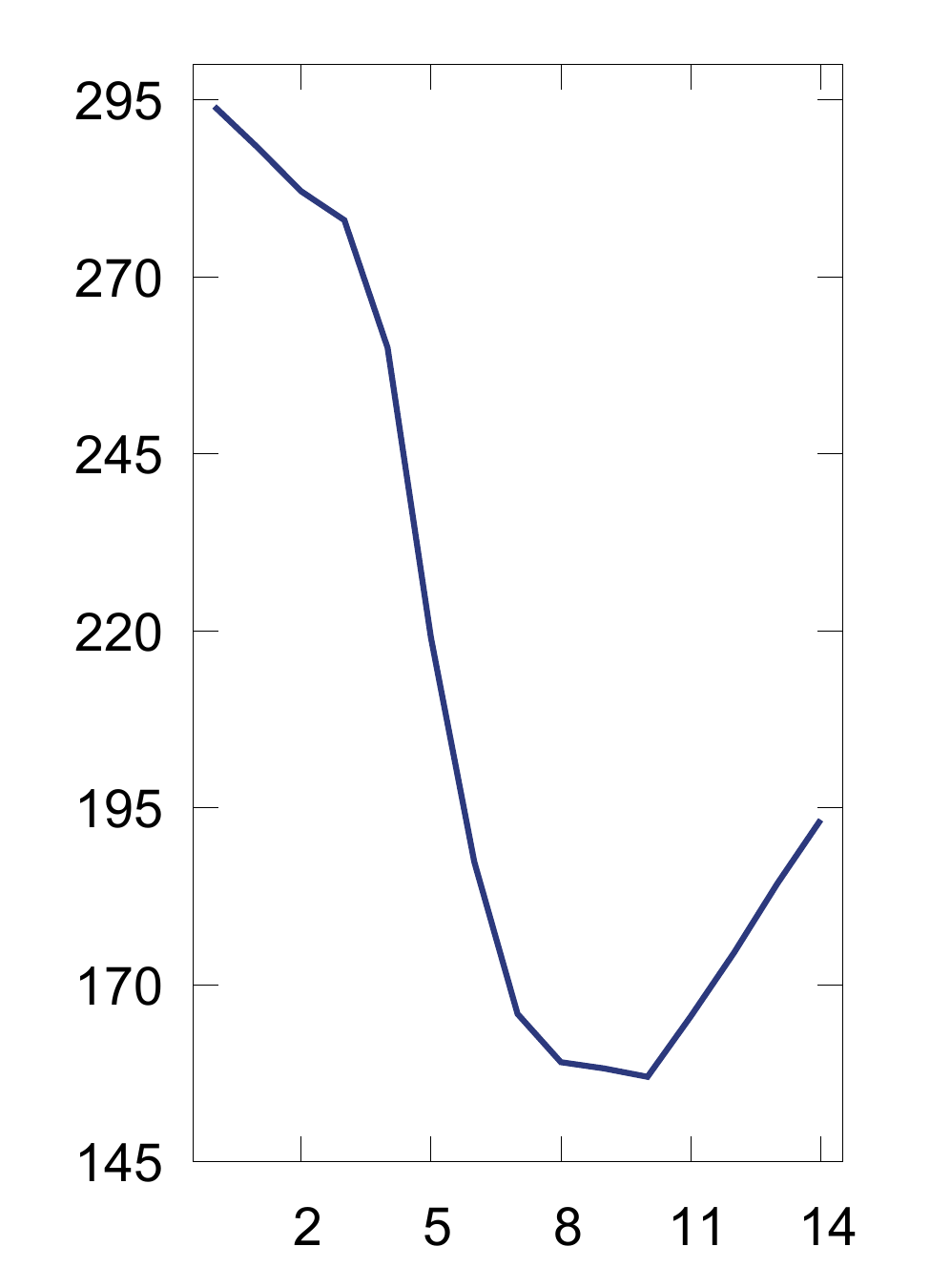}}\hspace{-1.1em}
\subfloat[$\lambda_{point}$]{\includegraphics[width=0.16\linewidth]{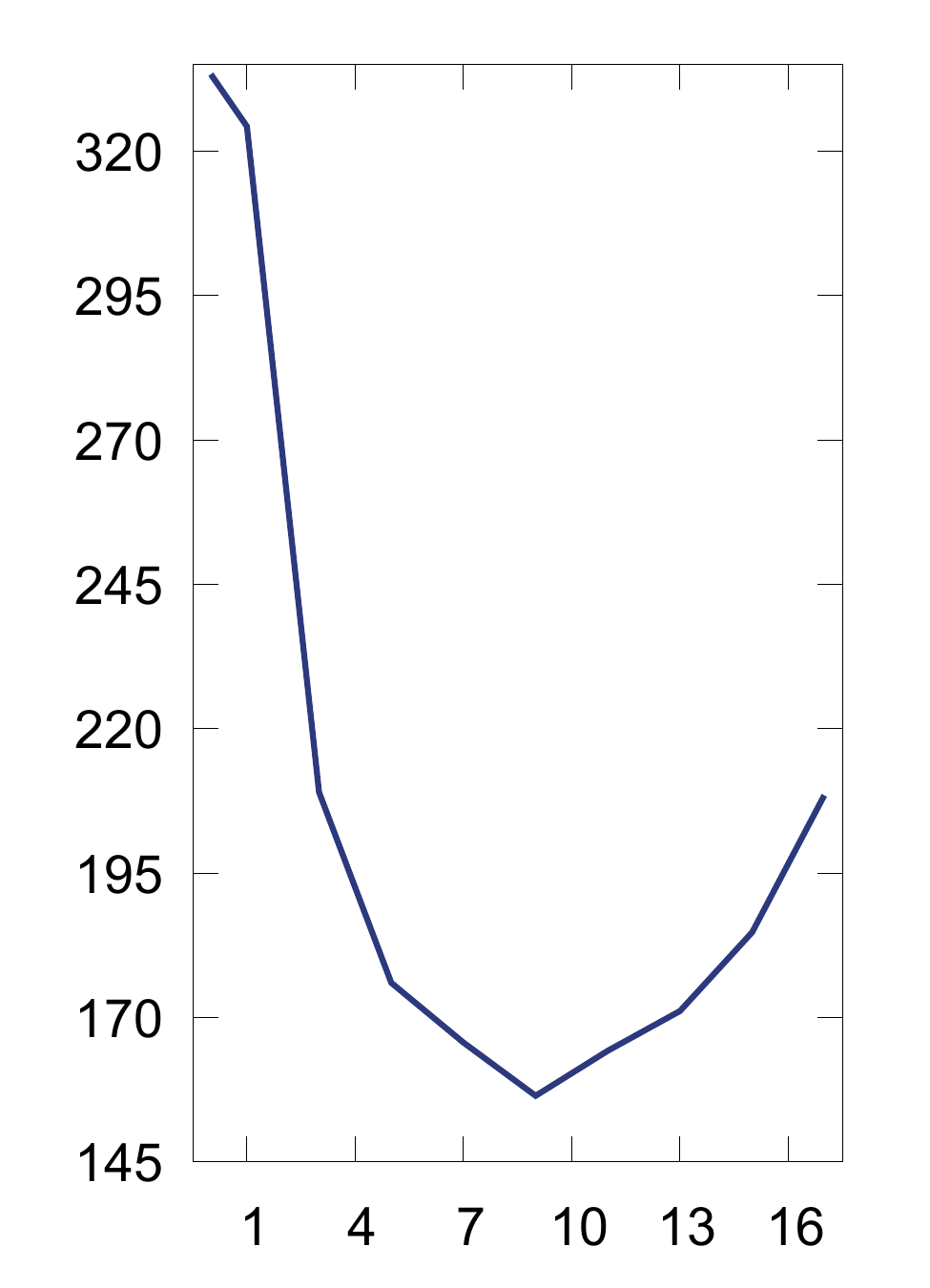}}\hspace{-1.1em}
\subfloat[$\lambda_{proj}$]{\includegraphics[width=0.16\linewidth]{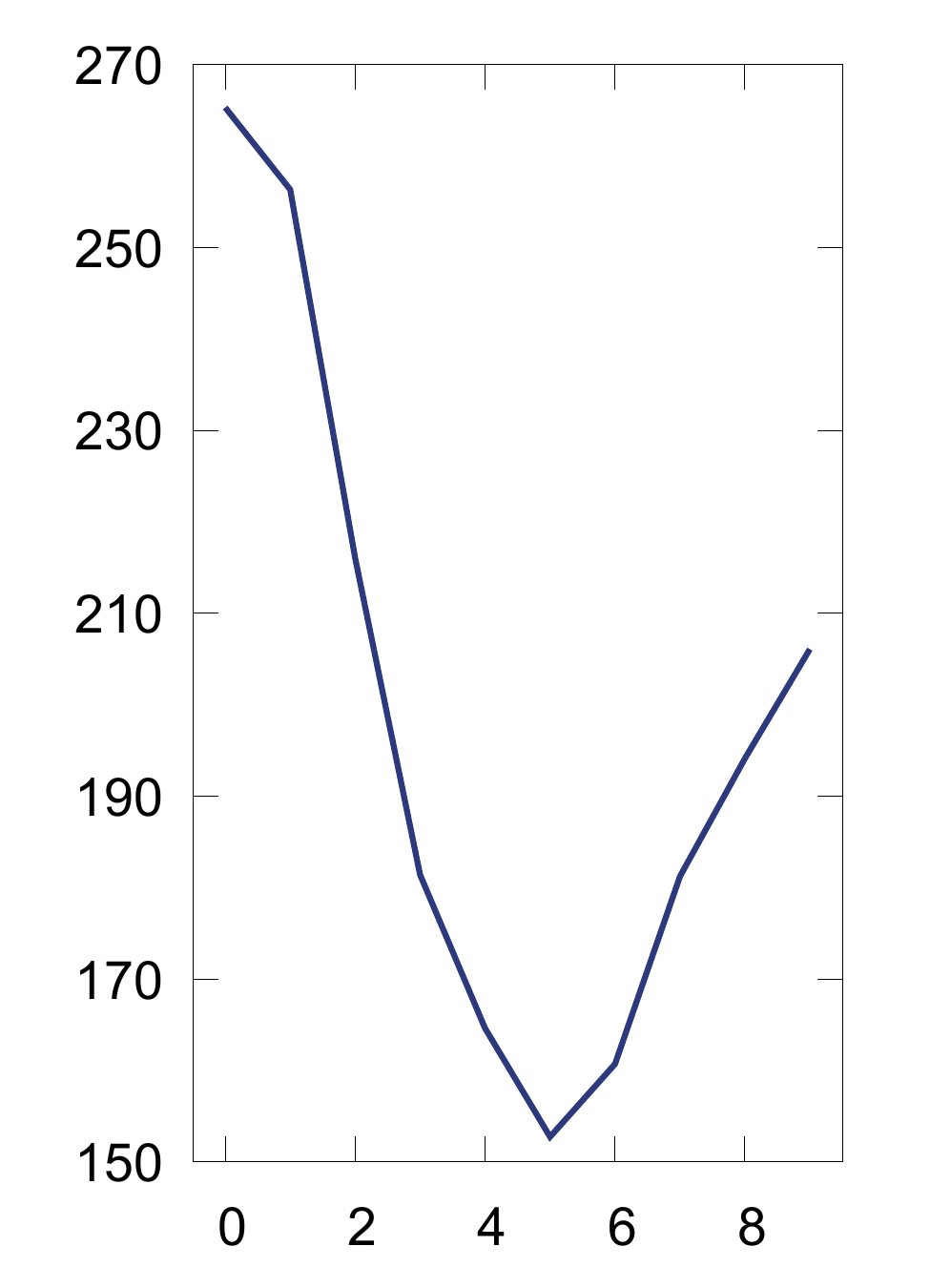}}\hspace{-1.1em}
\subfloat[$\lambda_{short}$]{\includegraphics[width=0.16\linewidth]{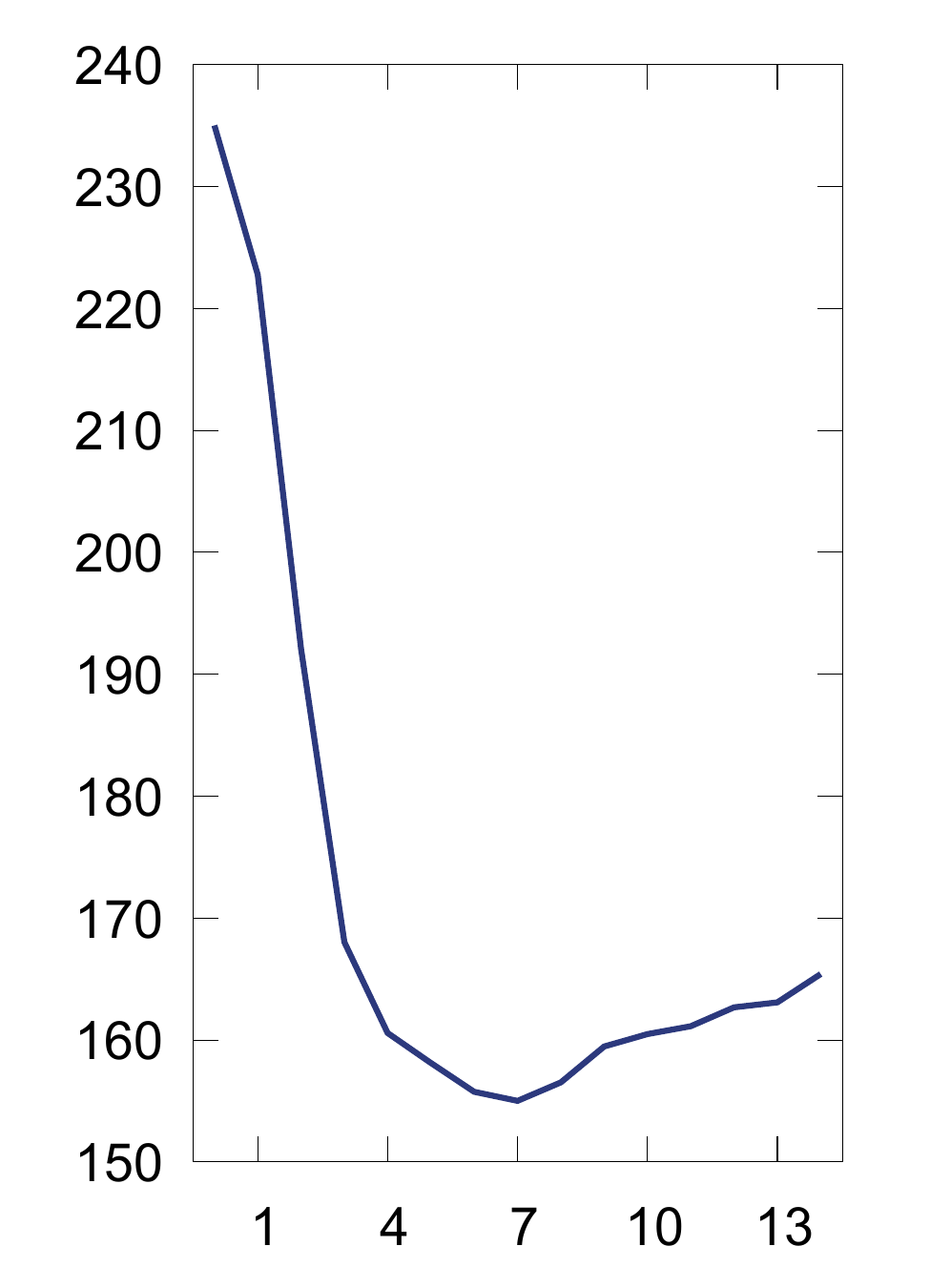}}
\caption{The root-mean-square error (\textbf{RMSE}) of the reconstructed depth maps against different parameter settings in all the sequences of the MPI Sintel~\cite{Butler:ECCV:2012} and Middlebury stereo datasets~\cite{scharstein2003high}. The fixed default values of the weight parameters were selected as follows: $\lambda_{rigid}{=}16$, $\lambda_{opti}{=}8$, $\lambda_{plane}{=}10$, $\lambda_{point}{=}9$, $\lambda_{proj}{=}5$ and $\lambda_{short}{=}7$.}
\label{fig:parameter}
\end{figure*}

\textbf{Parameters.}
We fine tune the optimal values of our important parameters via a quantitative analysis, as shown in Fig.~\ref{fig:parameter}. The x-axis is the range of each parameter, and the y-axis is the average accuracy for different input x values of all the sequences of the MPI Sintel dataset~\cite{Butler:ECCV:2012} and the Middlebury stereo dataset~\cite{scharstein2003high}. We used the fixed default parameter value in all the comparative experiments in Sec.~\ref{sec:results}.

\subsection{Quantitative Evaluation}\label{sec:quantitative_evaluation}

\begin{figure*}[!t]
\centering
\fontsize{8}{9}\selectfont
\begin{tabular}{p{0.1cm}p{3.0cm}p{3.0cm}p{3.0cm}p{3.0cm}p{3.0cm}}
\multicolumn{6}{c}{\includegraphics[width=1\linewidth]{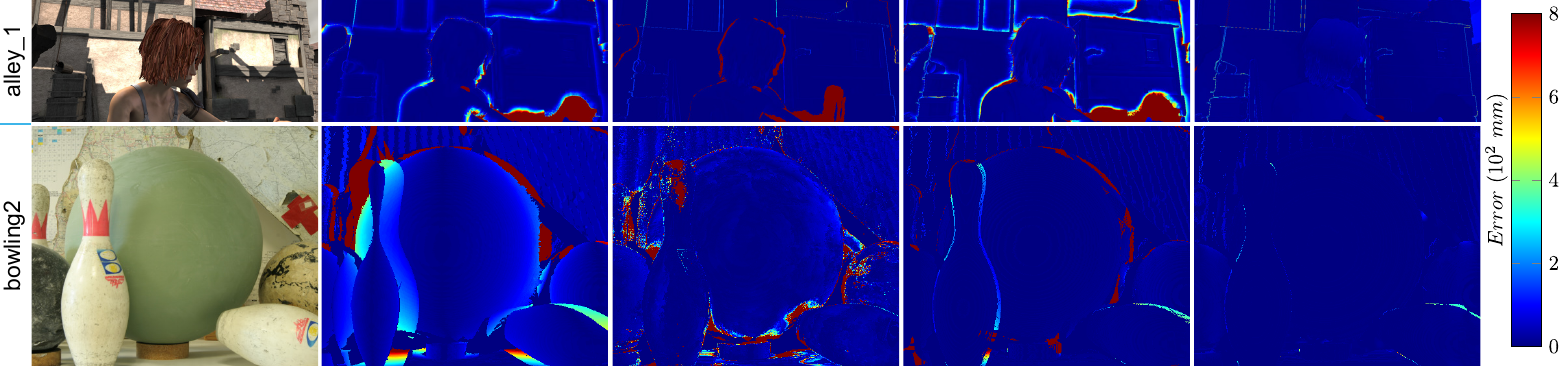}} \\
&(a) Color Images  & (b) Dolson et al.~\cite{Dolson10cvpr} & (c) Tracking-Based Method & (d) W/O occlusion detection and topology change detection & (e) Ours
\end{tabular}
\caption{Visualization of errors between reconstructed depth maps and ground truth. Our method produces more accurate depth maps.}
\label{fig:alley1error}
\end{figure*}

\textbf{Depth reconstruction.}
We compare our method with a tracking-based method and the method of Dolson et al.~\cite{Dolson10cvpr}. The tracking-based method exploits the motion information from depth maps with $E_{point}$, $ E_{plane}$, and $E_{rigid}$ and regularizes with $E_{reg}$ and $E_{short}$ but does not employ $E_{opti}$ or $E_{proj}$ to take advantage of the color images (Sec.~\ref{sec:optical_flow},~\ref{sec:spatialregu}). The post-process, solver (Sec.~\ref{sec:energy_minimization}) and the weights of individual terms are the same as those of our method. Meanwhile, we also make a comparison with the method by removing components for occlusion detection and topology change detection. In the hole filling post-process, the holes generated by occlusion are not filled. The state-of-the-art work~\cite{Dolson10cvpr} is, to the best of our knowledge, the only existing method for the temporal upsampling of depth maps given inputs similar to ours. Their method also uses the color information to interpolate the depth maps to recover more depth information. In the original datasets, each synthetic color image has its corresponding depth map. To evaluate the ability of reconstruction, we temporally downsample the depth frames, resulting in inputs similar to those from our hybrid camera. To evaluate the precision, we compute the \textbf{RMSE} of the reconstructed depth maps against the ground truth, as summarized in Tab.~\ref{tab:mpisinteldataset}. The results show that the reconstructed depth maps of our method are more accurate than those of others. This advantage can be seen based on the visualized errors in Fig.~\ref{fig:alley1error}.

\begin{table}[!h]
\caption{Quantitative evaluation of the depth reconstruction on the MPI Sintel dataset~\cite{Butler:ECCV:2012} and Middlebury stereo dataset~\cite{scharstein2003high}. As shown in this table, our method achieves the highest accuracy.}
\label{tab:mpisinteldataset}
\centering
\begin{tabular}{c|c|c|c|c}
\hline
$\mathbf{RMSE}$  & \parbox[][][c]{0.8cm}{Dolson et al.\protect\cite{Dolson10cvpr}} & \parbox[][][c]{1.8cm}{Tracking-Based Method} & \parbox[][][c]{1.8cm}{W/O occlusion detection and topology change detection} & \parbox[][][c]{0.8cm}{Ours} \\
\hline
$bamboo\_1$  & 614.3 & 784.4 & 263.33 & \textbf{238.3} \\
$alley\_1$  & 420.3 & 1048.5 & 403.05  & \textbf{61.8}\\
$sleeping\_2$  & 81.35 & 79.39 & 87.54 & \textbf{26.1}\\
$bandage\_1$   & 54.44 & 87.84 &  45.12 & \textbf{8.3}\\
$wood1$   & 159.29 & 273.85 & 88.23 & \textbf{81.93}\\
$bowling2$   & 482.82 & 433.17 & 174.23 & \textbf{155.92}\\
\hline
\end{tabular}
\end{table}

\textbf{Scene flow estimation.}
As discussed in Sec.~\ref{sec:sceneflowestim}, our joint optimization method can be used to estimate a scene flow given a set of input color images and corresponding depth maps captured at the same frame rate. We evaluate this scene flow method on the MPI Sintel dataset and Middlebury stereo dataset. Our approach is compared with the state-of-the-art techniques for scene flow estimation, including \emph{Layered-Flow}~\cite{sun2015cvpr}, \emph{SR-Flow}~\cite{QuirogaBDC14eccv} and \emph{PD-Flow}~\cite{jaimez2015icra}. The \textbf{RMSE}, average angular error (\textbf{AAE}), and end point error (\textbf{EPE}) are used as the error metrics~\cite{QuirogaBDC14eccv}. The quantitative evaluation results are given in Tab.~\ref{tab:sceneflow1}. The numerical results show that under all the error metrics and for almost all the sequences, our method outperforms the other scene flow methods and produces results closer to the ground truth.

\begin{table}[h]
\caption{Quantitative evaluation of scene flow estimation on the MPI Sintel dataset~\cite{Butler:ECCV:2012} and Middlebury stereo dataset~\cite{scharstein2003high}. The lower the error values, the better the performance.}
\label{tab:sceneflow1}
\centering
\begin{tabular}{c|c|c|c|c}
\hline
$\mathbf{EPE}$ & PD-Flow & SR-Flow & Layered-Flow & Ours \\
\hline
$alley\_1$ & 9.07 & 2.76 & 1.01 & \textbf{0.13} \\
$ambush\_5$ & 34.1 & 4.59 & 2.73 &  \textbf{0.84} \\
$cave\_2$ & 147.08 & 26.83 & 11.67 & \textbf{2.38} \\
$market\_2$ & 53.22 & 5.20 & 4.21 & \textbf{0.82} \\
$wood1$ & 49.87 & 9.72 & 10.09 & \textbf{0.50} \\
$bowling2$ & 109.60 & 10.82 & 5.84 &  \textbf{0.54}\\
\hline
$\mathbf{AAE}$ & PD-Flow & SR-Flow & Layered-Flow & Ours \\
\hline
$alley\_1$ & 1.53 & \textbf{0.98} & 2.10 & 1.85 \\
$ambush\_5$ & 1.43 & 1.19 & 1.11 & \textbf{0.74}  \\
$cave\_2$ & 1.60 & 1.44 & 1.57 & \textbf{1.30} \\
$market\_2$ & 0.99 & 0.75 & 1.24 & \textbf{0.28} \\
$wood1$ & 1.12 & 1.25 & 1.44 & \textbf{0.03} \\
$bowling2$ & 1.30 & 1.37 & 0.21 &  \textbf{0.02}\\
\hline
$\mathbf{RMSE}$ & PD-Flow & SR-Flow & Layered-Flow & Ours \\
\hline
$alley\_1$ & 10.56 & 4.03 & 3.39 & \textbf{0.52} \\
$ambush\_5$ & 71.65 & 5.98 & 5.21 & \textbf{2.01} \\
$cave\_2$ & 235.49 & 29.81 & 14.86 & \textbf{5.93} \\
$market\_2$ & 73.03 & 9.36 & 7.48 & \textbf{2.96} \\
$wood1$ & 125.44 & 10.09 & 30.20 & \textbf{0.77} \\
$bowling2$ & 356.65 & 10.75 & 18.56  & \textbf{1.86} \\
\hline
\end{tabular}
\end{table}

\subsection{Qualitative Evaluation}\label{sec:qualitative_evaluation}

\begin{figure*}[!t]
\centering
\includegraphics[width=1\linewidth]{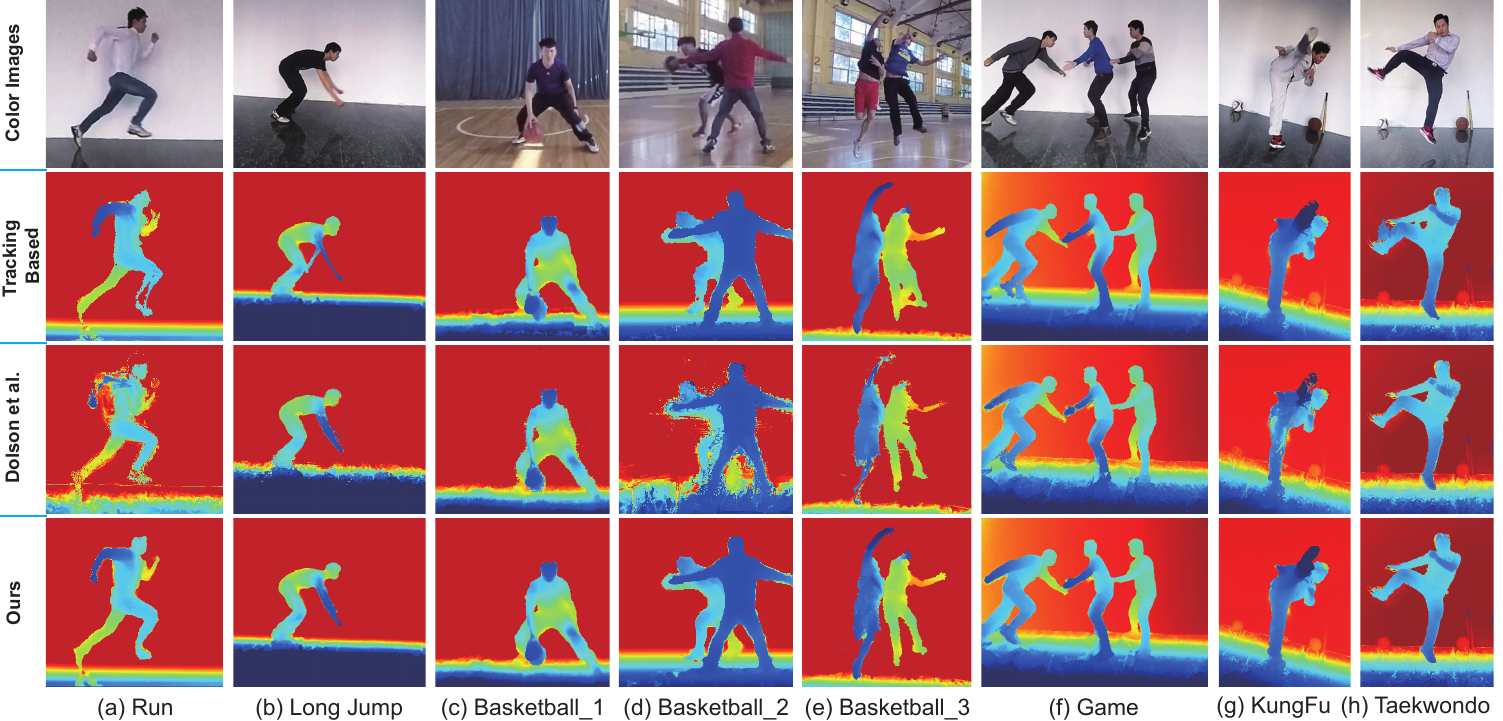}
\caption{Reconstructed depth maps of real-world scenes. From top to bottom, the rows present the following: input color images, depth maps reconstructed by the tracking-based method, results of the method of Dolson et al.\protect\cite{Dolson10cvpr}, and results of our method, respectively.}
\label{fig:realworldreault}
\end{figure*}

\begin{figure}[!t]
\centering
\subfloat[$I_1$]{\includegraphics[width=0.191\linewidth]{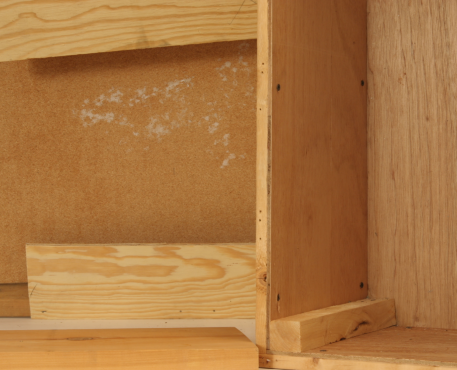}}\hspace{0.001em}
\subfloat[$I_2$]{\includegraphics[width=0.191\linewidth]{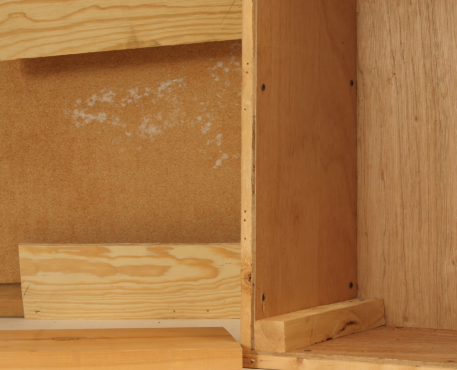}}\hspace{0.001em}
\subfloat[$I_3$]{\includegraphics[width=0.191\linewidth]{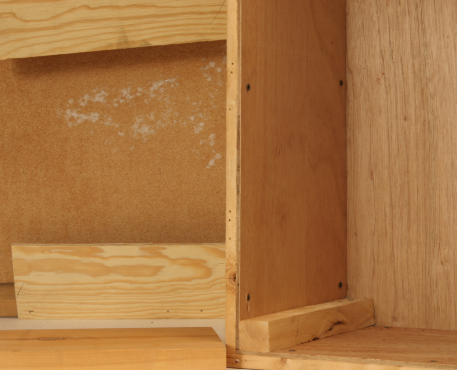}}\hspace{0.001em}
\subfloat[$I_4$]{\includegraphics[width=0.191\linewidth]{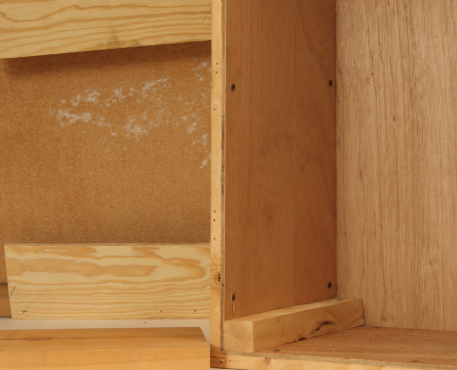}}\hspace{0.001em}
\subfloat[$I_5$]{\includegraphics[width=0.191\linewidth]{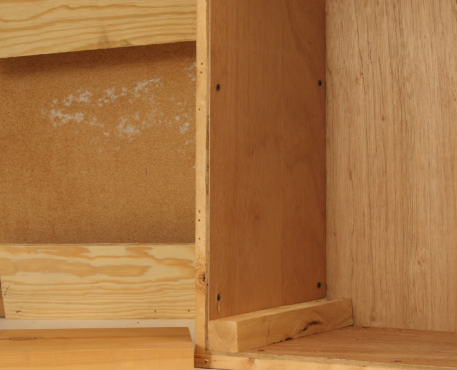}}\vspace{-1em}
\\
\subfloat[$D_1$]{\includegraphics[width=0.191\linewidth]{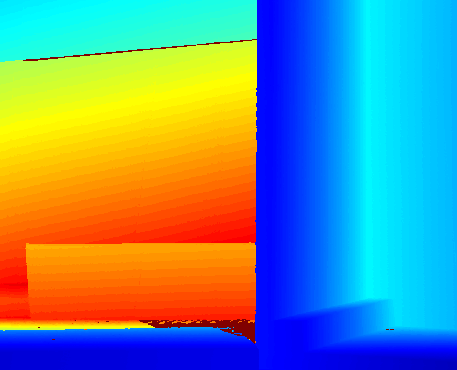}}\hspace{0.001em}
\subfloat[$D_5$]{\includegraphics[width=0.191\linewidth]{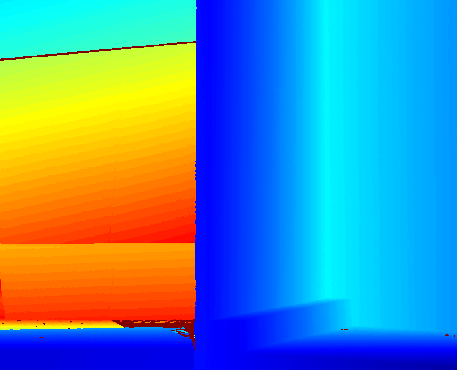}}\hspace{0.001em}
\subfloat[$D_3$(GT)]{\includegraphics[width=0.191\linewidth]{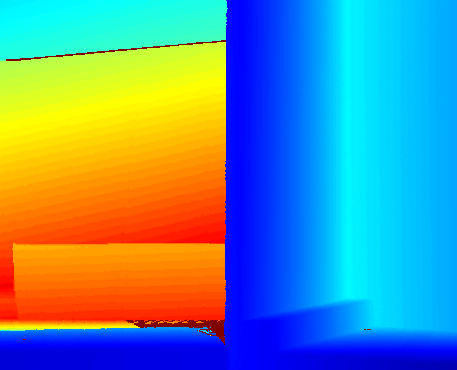}}\hspace{0.001em}
\subfloat[$D_3$(Ours)]{\includegraphics[width=0.191\linewidth]{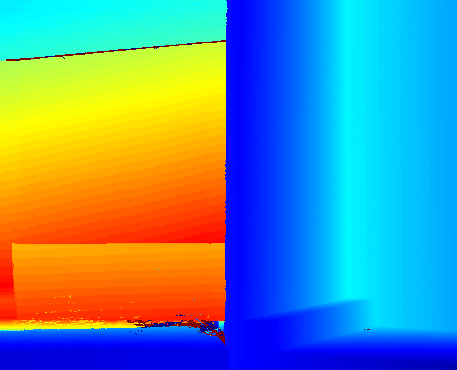}}\hspace{0.001em}
\subfloat[$D_3$(\cite{Dolson10cvpr})]{\includegraphics[width=0.191\linewidth]{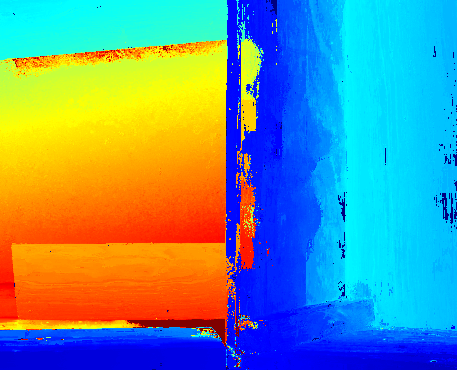}}
\caption{Results for the frames of sequence $wood1$ in the Middlebury stereo dataset~\cite{scharstein2003high}. (a)-(e) are input color images. (f) and (g) are input depth maps. (h) is the ground truth (GT) depth map corresponding to (c). (i) and (j) are the respective depth maps reconstructed by our method and \protect\cite{Dolson10cvpr} corresponding to (c).}
\label{fig:qualitexample}
\end{figure}

\begin{figure*}[!t]
\centering
\subfloat[Color Image]{\includegraphics[width=0.17\linewidth]{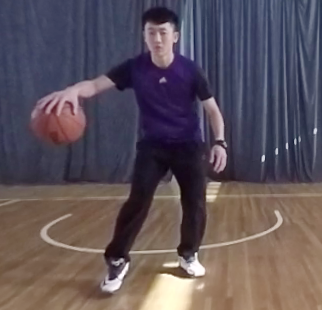}}\hspace{0.3em}
\subfloat[$1{:}8$]{\includegraphics[width=0.17\linewidth]{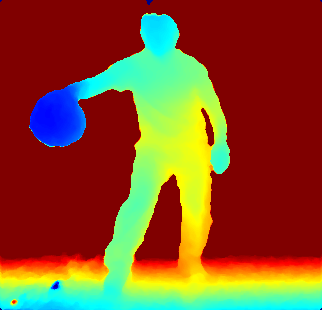}}\hspace{0.3em}
\subfloat[$1{:}16$]{\includegraphics[width=0.17\linewidth]{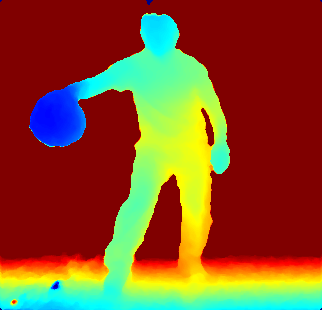}}\hspace{0.3em}
\subfloat[$1{:}24$]{\includegraphics[width=0.17\linewidth]{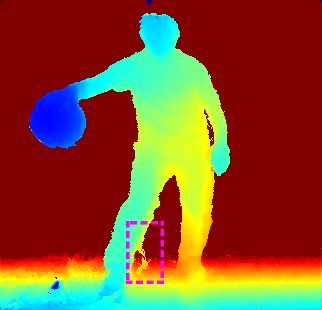}}\hspace{0.3em}
\subfloat[$1{:}32$]{\includegraphics[width=0.17\linewidth]{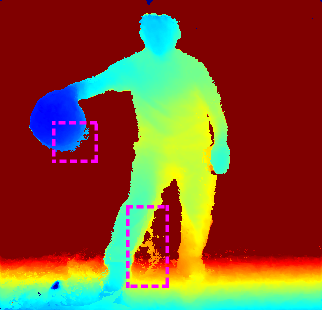}}
\caption{Downsampling results. (a) is the color image corresponding to the reconstructed depth map. (b) (c) (d) and (e) are reconstructed depth maps with frame-rate ratios of the depth maps generated by Kinect V2 to the color images taken by the GoPro of $1{:}8$, $1{:}16$, $1{:}24$, and $1{:}32$, respectively.}
\label{fig:downsampledepth}
\end{figure*}

In this subsection, we will evaluate our method on the data of real-world scenes and compare it to the technique of Dolson et al.~\cite{Dolson10cvpr} and the tracking-based method. We have made the comparisons on multiple challenging examples (see the accompanying video).  Fig.~\ref{fig:realworldreault} gives representative results. Since the tracking-based method does not take the color information into account, it may fail to reconstruct the depth information for a scene with fast motion. In contrast, our method employs the color information to evaluate the motion flow information, thus obtaining more accurate correspondence across frames.

The method of Dolson et al.~\cite{Dolson10cvpr} employs a d-dimensional Gaussian filtering framework to interpolate depth maps by encoding color, time, depth and location. However, their method does not take geometry into account, which is important to maintain the structures of objects. Without the constraint of this geometry prior, their method will easily choose incorrect reference depth data to interpolate intermediate depths. As shown in Fig.~\ref{fig:realworldreault}, their method causes artifacts with serious noise, as it patches the depth maps of foreground and background objects with the background, foreground or invalid data without the restraint of geometry. This problem is obvious in Fig.~\ref{fig:realworldreault}(d) and (e), as the gym is so deep (more than 6 meters) that the depth difference between the foreground and the background is very large. Our approach achieves much better results, mainly because of our careful consideration of the color (for optical flow), space and time information, as well as the topology and geometric relationships. As shown in another example in Fig.~\ref{fig:qualitexample}, when the colors of the foreground and background are similar, without geometry regularization, the approach of Dolson et al.~\cite{Dolson10cvpr} easily produces artifacts.

The frame-rate ratio of the depth maps generated by Kinect V2 to the color images taken by the GoPro is $1{:}8$. We also tested our method on $1{:}16$, $1{:}24$ and $1{:}32$ ratios, which are simulated by extracting the Kinect's depth maps at intervals of 2, 3 and 4 frames, respectively. It is expected and shown in Fig.~\ref{fig:downsampledepth} that a higher ratio will lead to more artifacts (e.g., on the legs in this example). These artifacts are mainly due to the errors of topology change detection caused by the cumulative error of optical flow.

\subsection{Impact of Individual Components}\label{sec:impeachterm}

\begin{figure*}[!t]
\centering
\subfloat[Color Image]{\includegraphics[width=0.16\linewidth]{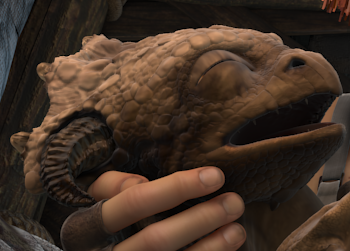}}\hspace{0.01em}
\subfloat[W/O $E_{point}$]{\includegraphics[width=0.16\linewidth]{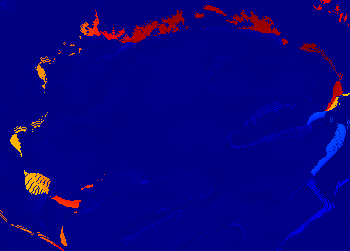}}\hspace{0.01em}
\subfloat[W/O $E_{plane}$]{\includegraphics[width=0.16\linewidth]{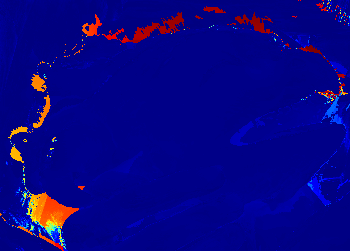}}\hspace{0.01em}
\subfloat[W/O $E_{opti}$ \& $E_{proj}$]{\includegraphics[width=0.16\linewidth]{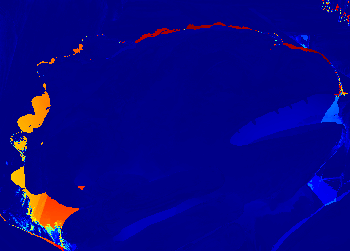}}\hspace{0.01em}
\subfloat[W/O $E_{rigid}$]{\includegraphics[width=0.16\linewidth]{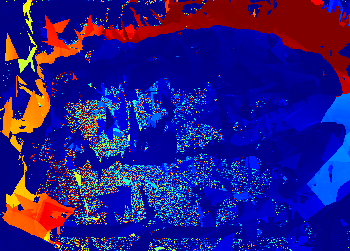}}\hspace{0.01em}
\subfloat[W/O $E_{reg}$]{\includegraphics[width=0.16\linewidth]{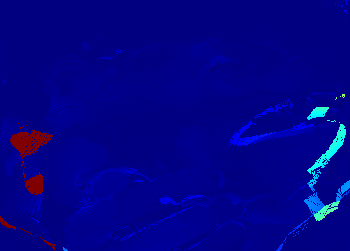}}\hspace{0.01em}
\\\vspace{-1em}
\subfloat[W/O $E_{short}$]{\includegraphics[width=0.16\linewidth]{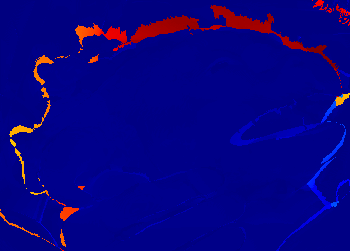}}\hspace{0.01em}
\subfloat[W/O Occlusion Detection]{\includegraphics[width=0.16\linewidth]{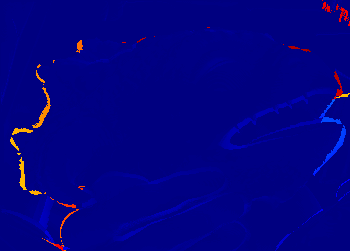}}\hspace{0.01em}
\subfloat[W/O Topology Change Detection]{\includegraphics[width=0.16\linewidth]{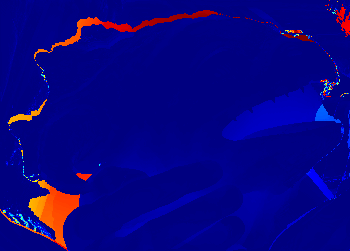}}\hspace{0.01em}
\subfloat[W/O Hole Filling]{\includegraphics[width=0.16\linewidth]{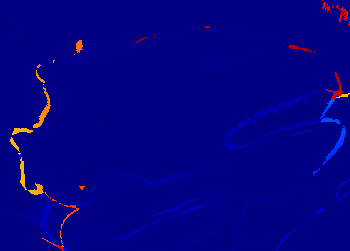}}\hspace{0.01em}
\subfloat[Ours]{\includegraphics[width=0.16\linewidth]{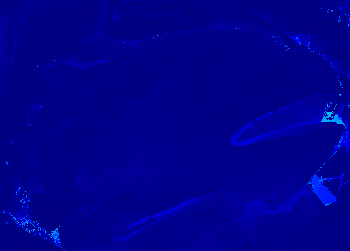}}\hspace{1.5em}
\subfloat{\includegraphics[width=0.06\linewidth]{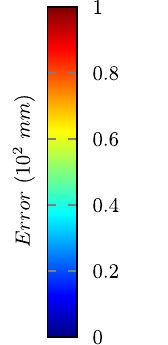}}
\caption{An evaluation of the importance of individual components. (a) is an input color image whose corresponding depth map needs to be reconstructed. (b)-(j) are the visualizations of error (the difference between the reconstructed depth maps and the ground truth) in our ablation study.}
\label{fig:compevalue}
\end{figure*}

\begin{table*}[!t]
\caption{Quantitative evaluation of component importance on the MPI Sintel dataset~\cite{Butler:ECCV:2012} and Middlebury stereo dataset~\cite{scharstein2003high}. Depth maps reconstructed by our full components method achieve the lowest \textbf{RMSE}.}
\label{tab:compevaluequantity}
\centering
\begin{tabular}{c|c|c|c|c|c|c|c|c|c|c}
\hline
$\mathbf{RMSE}$ & \parbox[][][c]{0.8cm}{W/O $E_{rigid}$} & \parbox[][][c]{1.3cm}{W/O $E_{opti}$ \& $E_{proj}$} & \parbox[][][c]{0.8cm}{W/O $E_{plane}$} & \parbox[][][c]{0.8cm}{W/O $E_{point}$} & \parbox[][][c]{0.8cm}{W/O $E_{reg}$} & \parbox[][][c]{0.8cm}{W/O $E_{short}$} & \parbox[][][c]{1.5cm}{W/O Occlusion Detection} & \parbox[][][c]{1.0cm}{W/O Hole Filling} & \parbox[][][c]{1.8cm}{W/O Topology Change Detection} & \parbox[][][c]{1.3cm}{With all the components}\\
\hline
$alley\_1$ & 901.46 & 784.4 & 429.72 & 379.45 & 340.14 & 315.84 & 323.41 & 316.46 & 253.59 & \textbf{238.3} \\
$wood1$ & 357.8 & 344.4 & 215.32 & 99.21 & 101.2 & 103.44 & 99.17 & 95.18 & 83.46 & \textbf{81.93} \\
$bowling2$ & 432.91 & 461.64 & 339.73 & 230.73 & 347.07 & 279.71 & 194.48 & 181.69 & 169.76 & \textbf{155.92} \\
\hline
\end{tabular}
\end{table*}

To evaluate the importance of each term, we conduct an ablation study of our framework. We evaluate using representative frames from the MPI Sintel dataset, which involve topology change, occlusion, and rigid and non-rigid transformation. In Fig.~\ref{fig:compevalue}, we visualize the errors against the ground truth when one component is removed. Moreover, we quantify the results using the \textbf{RMSE} between the reconstructed depth maps and ground truth, as shown in Tab.~\ref{tab:compevaluequantity}. Our method with all the components achieves the highest accuracy. It is shown that each component of our method is important, with $E_{rigid}$ being the most important term.

\textbf{Importance of $E_{point}$ and $E_{plane}$.}
These two terms take both corresponding point-pairs across the point clouds and the normal of a point cloud into account. Dropping either of them increases the reconstruction error, as shown in (b) and (c) of Fig.~\ref{fig:compevalue}.

\textbf{Importance of $E_{opti}$ and $E_{proj}$}.
$E_{opti}$ takes full advantage of 2D motion information from color images. $E_{proj}$ is used to connect 2D optical flow with 3D point cloud movement. As shown in Fig.~\ref{fig:compevalue}(d), when they are omitted, there are more artifacts in the reconstructed depth map. This is mainly because there are more mismatched nearest point relations in $E_{point}$ and $E_{plane}$ when the optical flow information is not used.

\textbf{Importance of $E_{rigid}$}.
This term is used to regularize the motion such that it is as locally rigid as possible. Since the local rigidity of the motion is very common in real-world scenes, this term plays a central role, as evidenced by the most serious errors in Fig.~\ref{fig:compevalue}(e).

\textbf{Importance of $E_{reg}$}.
$E_{reg}$ is employed to prevent the artifact generated in large-scale deformation~\cite{levi2015smooth}. The result without $E_{reg}$ is shown in Fig.~\ref{fig:compevalue}(f).

\textbf{Importance of $E_{short}$}.
This term is based on a temporal prior that enforces temporal smoothness and penalizes the jitter of a point cloud over time. Based on the observation that the speed of an object is essentially constant over a very short time, the $E_{short}$ term constrains the solution to a lower-dimensional subspace. The error against the ground truth is shown in Fig.~\ref{fig:compevalue}(g).

\textbf{Importance of Occlusion Detection}.
Occlusion will cause a mismatched correspondence relationship of $E_{proj}$, causing outliers of the optical flow and creating holes at the edge of the object. To disable the occlusion detection, we set $\mathbf{O}(\mathbf{v}_{s,i},\mathbf{C}_{t(k)+s})$ to 1 (Sec.~\ref{sec:enegy}) and do not fill the holes caused by occlusion. Fig.~\ref{fig:compevalue}(h) and (j) are almost the same, as the holes are generated by occlusion.

\textbf{Importance of Topology Change Detection}.
As shown in Fig.~\ref{fig:compevalue}(i), when topology changes are not considered, obvious artifacts occur on the horn and mouth of the dragon.

\textbf{Importance of Hole Filling}.
The hole-filling post-processing patches the invalid data in depth maps generated by occlusion and Kinect imaging. The dominant invalid depth data in the synthetic data is generated by occlusion. If we do not carry out hole filling, clear artifacts occur around the moving objects, as shown in Fig.~\ref{fig:compevalue}(j).

\subsection{Applications}
Various applications could benefit from our system with its reconstructed high-frame-rate depth maps. Here, we demonstrate three applications: fast human motion capture, rendering of stereoscopic images for a VR environment and the depth-of-field effect.

\begin{figure*}[!t]
\centering
\begin{tabular}{p{0.1cm}p{3.0cm}p{5.5cm}p{5.0cm}}
\multicolumn{4}{c}{\includegraphics[width=0.8\linewidth]{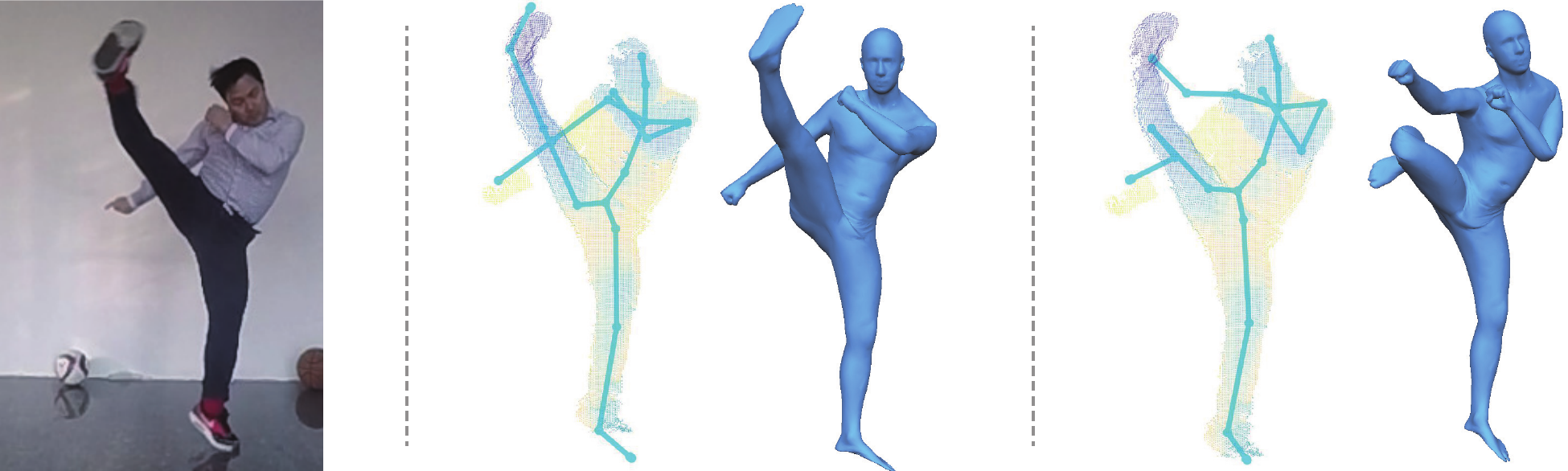}} \\
&(a) Color Images  & (b) With reconstructed high-frame-rate (240 FPS) depth maps & (c) With Kinect-captured (30 FPS) depth maps
\end{tabular}
\caption{Human motion capture. (a) is a color image taken by the GoPro. The corresponding point cloud and captured pose are shown in (b) and (c), respectively, which are captured from reconstructed high-frame-rate depth maps and Kinect-captured depth maps, respectively. The meshes in (b) and (c) are reconstructed via linear blending skinning (LBS) based on the currently captured 3D skeletal pose. The mesh in (b) is more reasonable than that in (c) and closer to the pose of the character in (a).}
\label{fig:humanmotiontracking}
\end{figure*}

\begin{figure}[!t]
\centering
\includegraphics[width=0.9\linewidth]{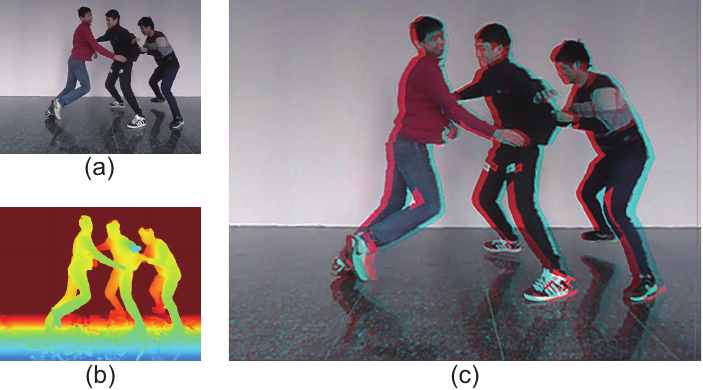}
\caption{Stereoscopic video rendering based on reconstructed depth maps and color images taken by the GoPro. (a) is the color image taken by the GoPro, (b) is the reconstructed depth map, and (c) is the stereoscopic image rendered by the method of de Albuquerque Azevedo et al.~\cite{de2014real}.}
\label{fig:stereoscopic}
\end{figure}

\textbf{Human Motion Capture.}
The state-of-the-art human motion capture studies include~\cite{wei2012accurate,chen2016realtime,liu2016kinect}. In those works, the input depth maps were generated by a Kinect at 30 FPS. To capture fast human motion, we make use of the depth maps from our system. We apply the full-body motion capture algorithm~\cite{wei2012accurate} to the depth maps from our hybrid system to capture the 3D skeletal poses of the fast human movement. As shown in Fig.~\ref{fig:humanmotiontracking}, thanks to the high frame rate and accurate depth maps of our hybrid system, the capturing algorithm performs better than when using the input from the Kinect directly.

\textbf{Stereoscopic Image Rendering.}
Stereoscopic videos provide more immersive experiences for virtual reality, such as 3DTV and head-mounted display. There are several methods that employ depth data to generate and edit stereoscopic images~\cite{mu2014stereoscopic,luo2015geometrically,du13tvcg}. In our system, a stereoscopic video can be easily acquired from the captured color images and reconstructed depth maps via a depth image-based rendering (DIBR)~\cite{de2014real}. DIBR is a technology that synthesizes virtual views of a scene using monocular color images and depth maps. With the help of DIBR, we synthesize a color video for the left eye and use the original color video for the right eye. Fig.~\ref{fig:stereoscopic} shows an example of the resulting stereoscopic images (the method of de Albuquerque Azevedo et al.~\cite{de2014real}). Please also find the rendered stereoscopic video in the accompanying video.

\begin{figure}[!t]
\centering
\subfloat[Color Image]{\includegraphics[width=0.30\linewidth]{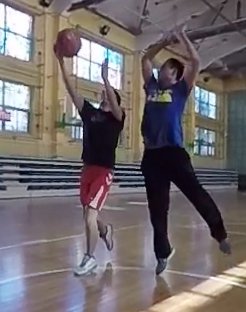}}\hspace{1em}
\subfloat[Focus on Players]{\includegraphics[width=0.30\linewidth]{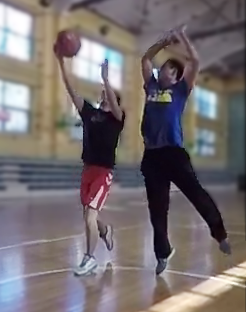}}\hspace{1em}
\subfloat[Focus on Background]{\includegraphics[width=0.30\linewidth]{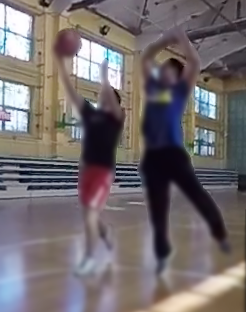}}
\caption{Depth-of-field effect: (a) is the entire sharp color image captured using the GoPro. To isolate a player from the background, (b) and (c) render with a small depth-of-field, causing the background and the players to be out of focus, respectively.}
\label{fig:depthoffield}
\end{figure}

\textbf{Depth-of-Field (DoF).} The DoF effect, which makes some objects in an image acceptably sharp and others blurry, is a common technique in photography used to emphasize a subject. Meanwhile, DoF is a render effect that is commonly applied to images or animation~\cite{lee2009real}. The usual method for producing a DoF effect in captured images is to use light field cameras, which capture images at a rate of three frames per second and are expensive. We use an image-based algorithm to render the DoF effect in RGB-D images. With the help of a reconstructed depth map, we can easily simulate the DoF effect and carry out refocusing in color images acquired with fast frame rates~\cite{moreno2007active}. We implement the method proposed by Kraus et al.~\cite{kraus2007depth} based on sub-images to change the DoF of the acquired images. As shown in Fig.~\ref{fig:depthoffield}, with the help of a reconstructed depth map, we render the DoF effect in GoPro-captured color images and refocus on players and the background separately to emphasize different subjects.

\subsection{Limitations and Future Work}
Our work has taken the first step in addressing the interesting issues of hybrid cameras in a temporal domain. Our technique can be improved in multiple aspects. First, our current unoptimized implementation is still too slow to support real-time performance capture. The bottleneck of our program is the transmission of data from the CPU to the GPU. We will completely implement the algorithm with CUDA to reduce the overhead of transfer and improve the throughput. Meanwhile, our framework reconstructs the sequence of depth maps together, using $\mathbf{D}_{k+1}$ to reconstruct previous depth frame $\mathbf{D}_{t(k)+s}$. The reconstructed result has a delay of $3\sim4$ milliseconds, even if the framework achieves real-time performance.

\begin{figure}[t]
\centering
\includegraphics[width=0.95\linewidth]{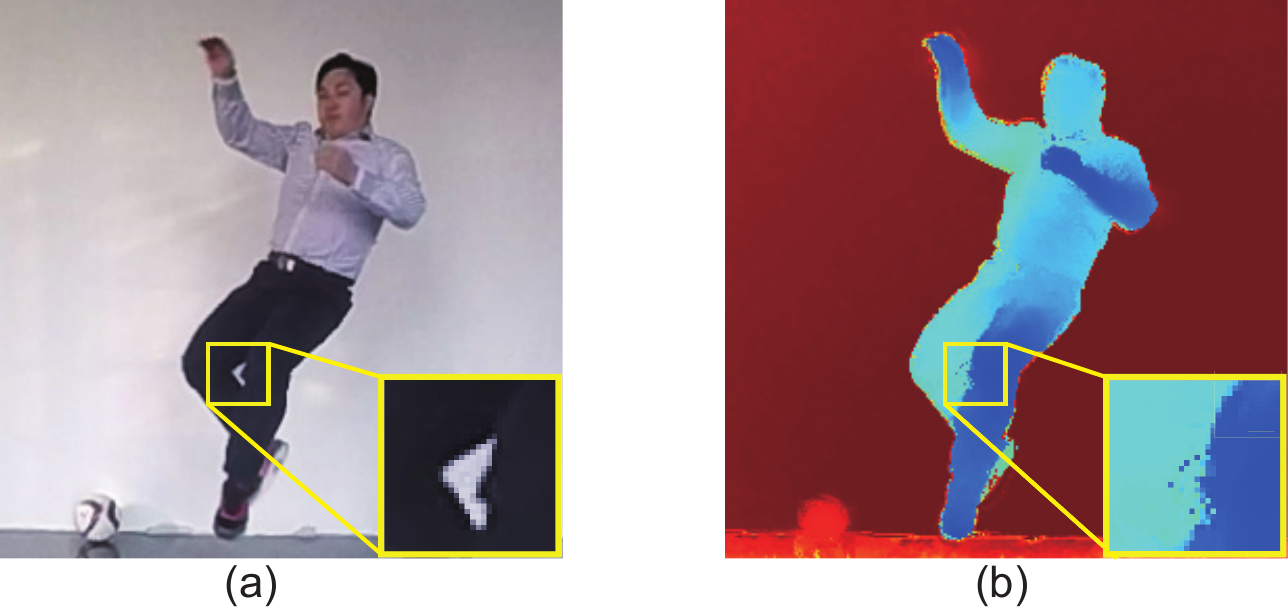}
\caption{Failure case. (a) is an input color image, and (b) is the corresponding reconstructed depth map. Our method fails to reconstruct depth data in the highlighted yellow box because of occlusion and fast motion.}
\label{fig:failurecase}
\end{figure}

Finally, our system does not reconstruct depth faithfully when motion is too fast and occlusion is too serious. As shown in Fig.~\ref{fig:failurecase}, in the gap between the two legs, where the occlusion is serious, and when the leg motion is very fast, our reconstruction result suffers from errors. The less accurate depth is due to a hole filling process error (Sec.~\ref{sec:holefilled}), as there is too little depth information regarding the gap. This problem can be mitigated by using more powerful hole filling methods, e.g., data-driven-based method~\cite{kwon2015data} and deep-learning-based method~\cite{liu2015deep}.


\end{document}